\documentclass[12pt,a4paper]{article}
\usepackage{amsmath}
\usepackage{amssymb} 
\begin{document}\begin{titlepage} 
\begin{flushright} IFUP--TH/2022\\
\end{flushright} ~\vskip .8truecm

~ 
\vskip 3.0truecm

\begin{center} 
  \Large\bf The continuation method and the real analyticity of the
  accessory parameters: the parabolic case
\end{center}
\vskip 1.2truecm
\begin{center}
{Pietro Menotti} \\ 
{\small\it Dipartimento di Fisica, Universit{\`a} di Pisa}\\ 
{\small\it 
Largo B. Pontecorvo 3, I-56127, Pisa, Italy}\\
{\small\it e-mail: pietro.menotti@unipi.it}\\ 
\end{center} 
\vskip 0.8truecm
\centerline{May 2022}
\vskip 1.2truecm
                                                              
\begin{abstract}
We give the proof of the real analyticity of the accessory parameters
in Liouville field theory as a function of the position of the sources
in the case in which in addition to elliptic sources, parabolic
sources are present. The method is a non trivial extension of the
elliptic case as it requires in an intermediate step the introduction
of a regulator. The treatment holds also in the case of the torus. A
discussion is given of the extension to higher genus surfaces.

\end{abstract}

\end{titlepage}

\section{Introduction}\label{introduction}

The accessory parameters play a very important role in Liouville
theory, which goes beyond the classical Riemann-Hilbert problem
\cite{bolibrukh} with implications in $2+1$ dimensional gravity
\cite{CMS2}, quantum conformal blocks \cite{FP,beccaria} and their classical
limit \cite{HJP,piatek,LLNZ,menottiConformalBlocks,alkalaev} through the
Polyakov relation \cite{CMS1,TZ}.

The nature of the dependence of these accessory parameters on the
position of the singularities and moduli is very important as it
intervenes in several major developments.

To this purpose we have the result of Kra \cite{kra} which in the case
of the sphere topology in presence of only parabolic and finite order
elliptic singularities proved using fuchsian mapping technique that
such a dependence is real analytic (not analytic). Finite order
singularities are those for which the source strength is given by
$\eta_k =(1-1/n)/2, n\in Z_+$ (see section \ref{lichtenstein}). However
such a technique is not applicable to the case of general elliptic
singularities which is the most common in applications.

The most direct method to determine the accessory parameters is the
monodromy matching condition in which one imposes the $SU(1,1)$ nature
of the monodromy around each singularities
\cite{keen,menottiAccessory, menottiHigherGenus}.  This procedure
leads to a system of implicit non linear equations. In the case when
only one independent parameter is present like the sphere topology
with four sources or the torus with one source, it was proven that
such accessory parameters are real analytic functions in the source
position or moduli space, almost everywhere (i.e. everywhere except
for a zero measure set)
\cite{menottiAccessory,menottiHigherGenus,menottiPreprint,menottiTorusBlocks}.
The qualification almost everywhere implies e.g. that we could not exclude
the occurrence of a number of cusps in the dependence of the accessory
parameters on the source positions, a phenomenon which may be expected
in the solution of a system of implicit equations.

This e.g. would put limits on the convergence radius of any power
series expansions of the parameters which go beyond the simple
position of the sources.

In a recent paper \cite{PMelliptic} it was proven that for any
collection of arbitrary elliptic singularities the dependence of the
accessory parameters is everywhere real analytic, exploiting a
completely different method i.e. the so called Le Roy-Poincar\'e
continuation method \cite{leroy,poincare,mahwin}, thus excluding the
appearance of non analyticity points.

In the present paper we give the extension of the continuation method
to the most general case i.e. the case where both arbitrary elliptic
and parabolic singularities are present.  This is not trivial: in fact the
largest part of the paper by Poincar\'e \cite{poincare}, which
is not interested in the analytic properties but simply in the
construction of the solutions by iterated power expansions, is devoted
to the treatment of parabolic sources, sometime called punctures or
singularities of the third type. Such problem cannot be treated by the
simpler method of Picard \cite{picard1,picard2}.

Due to the highly singular nature of the parabolic sources at an
intermediate step in the procedure of \cite{poincare}  a smooth, say
$C^\infty$, (non analytic) regulator is introduced. This may appear to
destroy the real analytic structure of the problem and how to overcome
such difficulty is one of the main subject of the present paper.

We work explicitly in the topology of the sphere and the final result
is that all the accessory parameters in the most general case when
both parabolic and arbitrary elliptic singularities are present, are
real analytic functions of the source positions everywhere except when
two source positions meet. This is the widest range of analyticity
which can be obtained.

The paper is structured as follows. In section \ref{lichtenstein} we
give the Lichtenstein decomposition of the conformal factor when in
addition to elliptic singularities also parabolic singularities
appear; this is similar but not exactly the same as the decomposition
when only elliptic singularities are present. In section
\ref{poincareprocedure} we summarize the construction of the solution
of the non linear equation. In section \ref{linearsection} we give the
solution of a linear partial differential equation which is the main
tool for solving the non linear Liouville equation. This is done first
in the case when only one parabolic singularity is present.  In
section \ref{twoparabolic} we give the extension of the procedure to
the case when more parabolic singularities are present.

Once the linear problem is solved the subsequent steps in the
procedure are similar to the pure elliptic case. In section
\ref{inheritance} we prove by an induction procedure that all the
functions which appear in the power expansion are analytic in the
position of the singularities and due to the uniform convergence this
property extend also to their sum.  Such analytic property is also
inherited by the solution of the original Liouville equation. In
section \ref{parametersanalyticity} we use the obtained results to
prove the real analytic dependence of the accessory parameters on the
positions of the sources.  With regard to the extension to higher
genus, the case of the torus presents no problem. The reason is that
we know the explicit Green function on the torus and thus its analytic
properties. For $g\geq2$ again this is bound to the analytic
properties of the Green function on higher genus surfaces about which
we do not possess explicit representations.  Probably some analytic
properties can be obtained by general considerations.  In order to
make the paper more readable we relegated in six Appendices the proof
of some results which are technical but fundamental for the
developments of the text.

The procedure of \cite{poincare} of writing the solution to the
Liouville equation in terms of uniformly convergent series is quite
powerful.  The result on the real analyticity of the conformal factor
is more general that the analyticity of the accessory parameters,
being the last fact a simple consequence of the previous.  The method
may be applied to other problems both in deriving qualitative
properties and quantitative bounds.

\section{The Lichtenstein decomposition in presence \break
  of parabolic sources}\label{lichtenstein}

The Liouville equation is
\begin{equation}\label{liouvilleequation}
\Delta\phi=e^\phi
\end{equation}
with the boundary conditions at the elliptic singularities
\begin{equation}
  \phi+2\eta_k \log|z-z_k|^2={\rm bounded},~~~~~~~\eta_k<\frac{1}{2}
\end{equation}
and at the parabolic singularities $z_P$ with the condition
\begin{equation}
\phi+\log|z-z_P|^2+\log\log^2|z-z_P|={\rm bounded}~
\end{equation}
and for sphere topology subject at infinity to the condition
\begin{equation}
  \phi+2\log|z|^2={\rm bounded}~~.
\end{equation}

The procedure \cite{lichtenstein} starts by constructing a positive
function $\beta$ everywhere smooth except at the elliptic sources
where it obeys the inequalities
\begin{equation}\label{inequality1}
0<\lambda_m<(|z-z_k|^2)^{2\eta_k}\beta(z) <\lambda_M
\end{equation}
and at the parabolic sources where we have
\begin{equation}\label{inequality2}
\beta=\frac{8}{|z-z_p|^2\log^2|z-z_P|^2}(1+ c(z)) 
\end{equation}
with
\begin{equation}
|c(z)|< \frac{c}{-\log|z-z_P|^2}
\end{equation}
in a neighborhood of $z_P$,
and for $|z|>\Omega$, being $\Omega$ the radius of a disk which
include all singularities $\beta$ is subject to the bounds
\begin{equation}\label{betainfinity}
0<\lambda_m<\beta |z|^4<\lambda_M ~. 
\end{equation}

It is possible to define a function $\nu$ such that in
$C\backslash\{z_k,z_P\}$
\begin{equation}\label{nuinfinity}
  \Delta\nu=\beta,~~~~\nu\approx -2\log|z|^2
  ~~~~{\rm for}~~z\rightarrow\infty~.
\end{equation}

First we write
\begin{equation}
\phi_1 = \sum_k(-2\eta_k)\log|z-z_k|^2-\sum_P\log|z-z_P|^2
\end{equation}
and then
\begin{equation}
\nu=\phi_1 +\frac{1}{4\pi}\int\log|z-z'|^2\beta(z')d^2z'~.
\end{equation}
Notice that also in presence of parabolic singularities $\int \beta(z)
d^2z<\infty$.

We must satisfy the sum rule
\begin{equation}\label{sumrule}
\frac{1}{4\pi}\int\beta(z')d^2z'=-2 +\sum_k 2\eta_k +\sum_P 1
\end{equation}
in order to have the correct behavior at infinity (\ref{nuinfinity}).

Apart from these requirements $\beta$ is free and due to the
uniqueness theorem \cite{lichtenstein,PMexistence} the final result
for the field $\phi$ does not depend on the specific choice of
$\beta$.

As we are interested in the analytic properties of the accessory
parameters and more generally of the conformal field, it is better to
start from a $\beta$ which is real-analytic except obviously at the
singularities $z_k$ and $z_P$. The choice of the $\beta$ is not
unique. In Appendix A we give a specific real analytic choice for
$\beta$ but we remark that in the developments of the papers the
explicit form of the $\beta$ will not be needed.

Given $\nu$ we define $u$ by
\begin{equation}
\phi = \nu+u~.
\end{equation}
With such a definition the Liouville equation becomes
\begin{equation}\label{liouville2}
\Delta u = e^\nu e^u-\beta \equiv \theta e^u-\beta~.
\end{equation}

The function $\nu$ and consequently the function $\theta$
contain $\beta$ in the form
\begin{equation}\label{basic}
  e^\nu = e ^{\phi_1+\frac{1}{4\pi}\int\log|z-z'|^2\beta(z')d^2z'}=
  \prod_k [(z-z_k)(\bar z-\bar z_k)]^{-2\eta_k}\prod_P [(z-z_P)(\bar
    z-\bar z_P)]^{-1} ~e^I~.
\end{equation}
We shall be interested in the dependence of the accessory parameters
on the position of a given source keeping the other fixed.  In the
following we shall consider as moving source a parabolic singularity
which is the most difficult case. Thus we shall be interested in the
dependence of the conformal factor when some $z_P$ varies in a small
domain $D_P$ around an initial value $z^0_P$. We shall choose for such
a domain the disk around $z^0_P$ whose radius is $1/4$ of the minimal
distance of $z^0_P$ from all the other singularities.  As we shall
keep all $z_k$ fixed and one $z_P$ moving we shall write
\begin{equation}
I(z,z_P)=\frac{1}{4\pi}\int\log|z-z'|^2\beta(z',z_P)d^2z'
\end{equation}
and $\nu(z,z_P)$ for $\nu$.

The analytic properties of $I(z,z_P)$ both in $z$ and $z_P$ are
derived in Appendix D by applying the general results of Appendix F.
$I(z,z_P)$ is real analytic in $z$ except at the singularities $z_k,z_P$
and everywhere finite except at the parabolic singularities. As far as
the dependence on $z_P$ is concerned, the $I(z,z_P)$ for $z\neq
z_k,~z\neq z_P$ is real analytic in the position of the moving
singularity for $z_P\in D_P$.  We recall that $\beta$ behaves at the
parabolic singularities $z_P$ like
\begin{equation}
\beta\approx\frac{8}{|z-z_P|^2\log^2|z-z_P|^2}~.
\end{equation}
This implies that $I$ behaves at the $z_P$ as
\begin{equation}
I=-\log\log|z-z_P|^2+ {\rm const}
\end{equation}
and thus we have for $\theta$
\begin{equation}
  \theta \equiv e^\nu e^I \sim \frac{1}{|z-z_P|^2 \log^2|z-z_P|^2}~.
\end{equation}
Given the sum rule (\ref{sumrule}) the asymptotic behavior of $I$ is
\begin{equation}
 I(z,z_P) \approx \log|z|^2 (2\sigma+N_P-2)
\end{equation}
where $\sigma = \sum_k\eta_k$ and from this we deduce that for $z_P\in
D_P$
\begin{equation}\label{r1rr2}
  0<r_1<\frac{\theta}{\beta}<r_2<\infty~.
\end{equation}

In the following we shall be interested in the real analytic
properties of the conformal factor $\phi$.  We recall that a real
analytic function can be defined as a function of two real variables
$x$ and $y$ which locally can be expanded in a convergent power series
\begin{equation}
  f(x+\delta x,y+\delta y) - f(x,y)= \sum_{m,n} a_{m,n} \delta x^m
  \delta y^n~.
\end{equation}
Equivalently it can be defined as the value assumed by an analytic
function $f(z,z^c)$ of two complex variables when $z^c$ assumes the
value $\bar z$. In this paper as in \cite{PMelliptic} we shall adopt
the second setting and use for $f(z,\bar z$) the simplified notation
$f(z)$ with the definitions $\frac{\partial}{\partial
  z}=\frac{1}{2}(\frac{\partial}{\partial x}
-i\frac{\partial}{\partial y})$ and $\frac{\partial}{\partial \bar
  z}=\frac{1}{2}(\frac{\partial}{\partial x}
+i\frac{\partial}{\partial y})$.

\section{The equation $\Delta u =\theta e^u-\beta$}
\label{poincareprocedure}

As in the case of elliptic singularities the solution of the equation
\begin{equation}\label{originaleq}
\Delta u= \theta e^u-\beta~~~~~{\rm with}~~\theta=e^\nu\equiv r\beta 
\end{equation}
is reduced to the solution of a system of inhomogeneous linear
equation \cite{poincare}. This first step does non differ from the
case of only elliptic singularities but for the benefit of the
reader we give a brief summary here below; for full details see
\cite{poincare,PMelliptic}

We have as consequence of the sum rule (\ref{sumrule}) $0<r_1<r<r_2<\infty$
for certain $r_1,r_2$.

Let $\alpha$ be the minimum
\begin{equation}
\alpha=\min\bigg(\frac{\beta}{\theta}\bigg)=\frac{1}{\max r}
\end{equation}
which due to (\ref{r1rr2}) is a positive number.
Then we can rewrite the equation as
\begin{equation}\label{rewritten}
\Delta u = \theta e^u - \alpha\theta -\beta(1-\alpha r)~.
\end{equation}
As a consequence of the choice for $\alpha$ we have $\psi\equiv
\beta(1-\alpha r)\geq 0$.

Convert the previous equation to 
\begin{equation}\label{lambdaeq}
  \Delta u = \theta e^u-\alpha\theta -\lambda\psi
\end{equation}
and write
\begin{equation}\label{nonlinearseries}
  u=u_0+\lambda u_1+\lambda^2 u_2+\dots~.
\end{equation}  
We have to solve the system 
\begin{eqnarray}\label{nonlinearsystem}
  &&\Delta u_0 = \theta (e^{u_0}-\alpha)\nonumber\\
  &&\Delta u_1 = \theta e^{u_0}u_1-\psi\nonumber\\
  &&\Delta u_2 = \theta e^{u_0}(u_2+w_2)\nonumber\\
  &&\Delta u_3 = \theta e^{u_0}(u_3+w_3)\nonumber\\
  &&\dots
\end{eqnarray}  
where
\begin{equation}
  w_2 = \frac{u_1^2}{2},~~~~w_3 = \frac{u_1^3}{6}+u_1u_2,~~~~
  w_4 = \frac{u_1^4}{24}+\frac{u_1^2 u_2+u_2^2}{2}+u_1u_3,~~~~\dots
\end{equation}  
are all polynomials with positive coefficients. We see that in
the $n$-th equation the $w_n$ is given in terms of $u_k$ with $k<n$
and thus each of the equations (\ref{nonlinearsystem}) is a linear
equation.

The first equation is solved by $u_0=\log\alpha$ and from the
properties of the Laplacian $\Delta$ and eq.(\ref{nonlinearsystem}) we
have
\begin{eqnarray}\label{inequalities}
  &&|u_1| \leq \max\bigg(\frac{\psi}{e^{u_0}\theta}\bigg)\nonumber\\
   &&|u_2| \leq \max~|w_2|\nonumber\\
  &&|u_3| \leq \max~|w_3|\nonumber\\
  && \dots
\end{eqnarray}

Using the above inequalities one proves \cite{poincare,PMelliptic} that the
series (\ref{nonlinearseries}) converges for
\begin{equation}\label{lambda0bound}
  |\lambda|<\frac{\alpha(\log 4-1)}
  {\max|\frac{\psi}{\theta}|}
\end{equation}
and such convergence is uniform.

Thus we are able to solve the equation
\begin{equation}
\Delta u = \theta e^u - \alpha\theta -\lambda_0\psi
\end{equation}
for
\begin{equation}
0<\lambda_0<\frac{\alpha(\log4-1)}{\max~|\frac{\psi}{\theta}|}~.
\end{equation}
If $\lambda_0$ can be taken equal to $1$ the problem is solved.
Otherwise one can extend the region of solubility of our equation by
solving the equation
\begin{equation}\label{secondstep}
  \Delta u = \theta e^u - \theta \alpha-
    \lambda_0\psi
  -\lambda\psi\equiv
  \theta e^u - \varphi-\lambda\psi~.
\end{equation}

It is easily proven that the series in $\lambda$ converges with the
same radius given by (\ref{lambda0bound}) and thus in a finite number
of steps we reach the solution of eq.(\ref{lambdaeq}) for $\lambda=1$
i.e. the solution of (\ref{originaleq}).

Thus we see that the solution of the Liouville equation is reduced to
the solution of linear inhomogeneous equations of the form
\begin{equation}
\Delta u = \eta u-\beta
\end{equation}  
with $\eta$ equivalent to $\theta$ i.e.
$0<c_1<\frac{\eta}{\theta}<c_2$ \cite{poincare}. In the sequel we
shall encounter several times functions equivalent to $\theta$ and we
shall call them generically $\eta$.

\section{The equation $\Delta u = \eta u -\beta$}
\label{linearsection}

We saw in the previous section that the solution of the Liouville
equation has been reduced to the solutions of linear inhomogeneous
partial differential equations of the type $\Delta u = \eta u -\beta$
where $\eta$ has the same singularities as $\theta$ i.e.
$0<c_1<\frac{\eta}{\theta}<c_2<\infty$.

The main difference with the pure elliptic case is that in presence of
only elliptic singularities we have that
\begin{equation}\label{simpleintegral}
\frac{1}{4\pi}\int \log|z-z'|^2\beta(z') d^2z'
\end{equation}  
is finite for all $z$. Instead in presence of parabolic singularities
even though we have still
\begin{equation}
\int \beta(z') d^2z' <\infty
\end{equation}  
the integral (\ref{simpleintegral}) diverges at $z=z_P$. This makes
already the first step in the expansion procedure to fail.

This first difficulty is circumvented by noticing the
$\eta$ has at the parabolic points the same singularities of type
$1/(|z-z_P|^2\log^2|z-z_P|^2)$ as $\beta$ and thus by writing
\begin{equation}\label{subtraction}
u=u'+v
\end{equation}
where $v$ is bounded real analytic function we have
\begin{equation}\label{beta1equation}
  \Delta u'= \eta u' - \Delta v +\eta v -\beta\equiv
  \eta u' -\beta_1
\end{equation}
and by properly choosing the values of $v$ at the parabolic
singularities we can cancel the leading singularity of $\beta$ at the
parabolic points thus reaching a source $\beta_1$
\begin{equation}
  \beta_1\sim \frac{c}{|z-z_P|^2 \log^3|z-z_P|^2}~.
\end{equation}
We shall call such a singularity a tamed parabolic singularity.  For
an explicit real analytic choice of $v$ see Appendix A. We point out
that a bounded solution to eq.(\ref{beta1equation}) necessarily vanish
at all parabolic singularities. In fact if we have $u'(z_P)=c\neq 0$
then the general solution to (\ref{beta1equation})
\begin{equation}
\frac{1}{4\pi}\int\log|z-z'|^2 (\eta u'-\beta_1)d^2z' + {\rm const}
\end{equation}
is going to diverge at $z=z_P$ due to the parabolic divergence of
$\eta$.

We notice that with this subtraction procedure the parabolic
singularities of the source $\beta$ have been tamed: not so for the
$\eta$ which still contains the parabolic singularity at the $z_P$.
This difficulty is circumvented by first solving exactly around $z_P$
an approximate form of eq.(\ref{beta1equation}) where
$\eta$ is replaced by a simpler term.

We shall start from the case in which in addition to elliptic
singularities we have a single parabolic one and then give the
extension to more than one parabolic singularity.

We shall write $\eta=\eta_0+\eta'$ with, using $\zeta=z-z_P$,
\begin{equation}
\eta_0(\zeta)=\eta_E(\zeta)
\rho(\zeta)+\eta(\zeta,z_P)(1-\rho(\zeta));~~~~~~~~\eta'(\zeta,z_P) =
\rho(\zeta)(\eta(\zeta,z_P)-\eta_E(\zeta))
\end{equation}
and
\begin{equation}
  \eta_E = \frac{t}{|\zeta|^2 \log^2|\zeta|^2}
\end{equation}
being $t$ the coefficient of the parabolic singularity of $\eta$ at
$z_P$.  The $\rho(x)$ is a $C^\infty$ function which equals $1$ for
$0<|x|<R'<R$, decreases to zero for $R'<|x|<R$ and is zero for
$x>R$. We shall choose $R$ less than $1/4$ of the minimal distance of
$z_P$ from the nearest singularity $z_k$ and also $R<1/\sqrt{e}$.
Such a function is necessarily non real analytic but it can be easily
chosen as to restrict the non analyticity region to the two circles
$|z-z_P|=R'$ and $|z-z_P|=R$.  It is close to the generic regulator
adopted in \cite{poincare}.  This may appear to spoil the analyticity
of the treatment but when discussing the analytic properties of the
solution we shall see that exploiting the freedom in choosing $R'$ and
$R$ and the uniqueness theorem \cite{lichtenstein,PMexistence} we
shall obtain analyticity of the conformal factor everywhere except at
the singularities $z_k,z_P$. It is also possible to introduce a
regulator which is real analytic everywhere. This is given is Appendix
E and it would keep the treatment within the realm of real analytic
functions. However both in the present section and in section
\ref{twoparabolic} we shall adopt the above described $C^\infty$
regulator, as it is simpler.

We set
\begin{equation}\label{mdefinition}
m=\max\bigg(\frac{\eta_0}{\eta}\bigg)
\end{equation}
which is a finite number and for $R$ small enough we have
\begin{equation}\label{etaprimeovereta0}
\max\bigg|\frac{\eta'}{\eta_0}\bigg|<1~.
\end{equation}
 This inequality
will be relevant at the end of the present section.

Thus we shall first solve the simpler equation
\begin{equation}\label{simplerequation}
  \Delta u = \lambda \eta_0 u -\beta_1=
  \lambda (\eta_E\rho +\eta(1-\rho)) u -\beta_1
\end{equation}
where not to overburden the notation we reverted to the notation $u$
for $u'$. Eq.(\ref{simplerequation}) in the neighborhood of $z_P$,
$|\zeta|<R'$ reduces exactly to
\begin{equation}
  \Delta u = \lambda \eta_E u -\beta_1~.
 \end{equation}
It is immediately seen that $s=\log^h\frac{1}{|\zeta|^2}$ solves the equation
\begin{equation}\label{}
\Delta \log^h\frac{1}{|\zeta|^2} = \lambda\eta_E\log^h\frac{1}{|\zeta|^2}
\end{equation}
with $\lambda=h(h-1)/t$ which for $h<0$ covers all possible values
of $\lambda>0$.

Thus we introduce the function $\nu$ given by
\begin{equation}
  \nu = -\log|\zeta|^2\rho(\zeta)+1-\rho(\zeta) =
  1+(-\log|\zeta|^2-1)\rho(\zeta)\geq1
\end{equation}
as the support of $\rho$ is less than $1/\sqrt{e}$, and thus $\nu^h$
provides an exact solution of
\begin{equation}\label{}
\Delta \nu^h = \lambda\eta_0 \nu^h
\end{equation}
for $|\zeta|<R'$ where $\eta_0=\eta_E$.

Writing $u=u'+\mu \nu^h$ with $\mu$ a constant, we obtain 
\begin{equation}\label{uprimeeqaution}
\Delta u'=\lambda\eta_0u'-\beta_1+\mu\psi
\end{equation}
\begin{equation}
\frac{4h(h-1)}{t}= \lambda,~~~~~~~~h<0
\end{equation}
where
\begin{equation}
\psi=\lambda\eta_0\nu^h-\Delta\nu^h~.
\end{equation}
The $\psi$ vanishes for $|\zeta|<R'$ and equals $\lambda\eta_0=
\lambda\eta$ for $|\zeta|>R$.
Notice that $\lambda$ is present in $\psi$.
Moreover
$\int\psi d^2z = \lambda\int\eta_0\nu^hd^2z>0$~.

Keeping $\lambda$ fixed we
shall solve
\begin{equation}\label{masterequation}
\Delta u'=\lambda_1\eta_0u'-\beta_1+\mu(\lambda_1)\psi
\end{equation}
expanding in series in $\lambda_1$.
We shall show that for $\lambda$ sufficiently small i.e. $|\lambda|<
\lambda_0$ such an expansion in $\lambda_1$ converges uniformly for
$\lambda_1=\lambda$ thus providing a solution to (\ref{uprimeeqaution})
for a proper value of $\mu$.

\bigskip

To zero order we have
\begin{equation}\label{zeroorder}
\Delta u'_0 =-\beta_1+\mu_0\psi
\end{equation}
with the condition 
\begin{equation}
-\int \beta_1 d^2z+\mu_0\int \psi d^2z=0
\end{equation}
as to comply with the finiteness of $u'_0$ at infinity. Due to the
vanishing of $u'$ at the parabolic singularity the solution to
(\ref{zeroorder}) is given by
\begin{equation}
u'_0=\int G_1(z,z')\big(-\beta_1(z')+\mu_0\psi(z')\big) d^2z'
\end{equation}
where $G_1$ is the subtracted Green function
\begin{equation}
G_1(z,z')=\frac{1}{4\pi}\log\bigg|\frac{z-z'}{z_P-z'}\bigg|^2~.
\end{equation}
Moreover due to the bound
proved in Appendix C we have
\begin{equation}\label{secondPbound}
|u'_0|< \frac{{\rm const}}{L(z-z_P)}
\end{equation}
with
\begin{equation}\label{L}
L(z-z_P) =\log\bigg(\frac{1+2|z-z_P|^2}{|z-z_P|^2}\bigg)>\log 2.
\end{equation}
To first order we have 
\begin{equation}
  \Delta u'_1 =\eta_0 u'_0+\mu_1\psi
\end{equation}
with the condition
\begin{equation}
  \int\eta_0 u'_0 d^2z+\mu_1\int\psi d^2z=0~.
\end{equation}

It is useful to define a function $\zeta_1$ which is positive, never
vanishing with the same asymptotic behavior and singularities of $\eta$
except at $z_P$ where it as the tamed singularity
$1/(|z-z_P|^2\log^3|z-z_P|^2)$. It is defined by
\begin{equation}\label{zeta1}
\zeta_1=\frac{\eta}{Z L}
\end{equation}
with $L$ given by eq.(\ref{L}) and
\begin{equation}
Z=\prod_s \frac{(|z-z_s|^2)^{-2\eta_s}}{(1+|z-z_s|^2)^{-2\eta_s}}
\end{equation}
being $z_s$ the positions of the elliptic singularities with $\eta_s<0$.
Then we can write
\begin{equation}
  \Delta u'_1 =\eta_0 u'_0+\mu_1\psi\equiv\zeta_1\varphi_1
\end{equation}
with a bounded $\varphi_1$.
To higher orders
\begin{equation}\label{recursiveeq}
  \Delta u'_k =\eta_0 u'_{k-1}+\mu_k\psi\equiv\zeta_1\varphi_k
\end{equation}
\begin{equation}\label{mueq}
  \int\eta_0 u'_{k-1} d^2z+\mu_k\int\psi d^2z=0~.
\end{equation}
We have 
\begin{equation}
  \int\psi d^2z= \lambda\int\eta_0 \nu^hd^2z>\lambda\int_{R_c}\eta_0 d^2z>0
\end{equation}
being $R_c$ is the region outside the circle of radius $R$, where $\nu=1$.

\bigskip

The instrumental bounds in the following are
\begin{equation}\label{Cinequality}
  |u'_k|<\frac{b_2}{L(z-z_P)} \gamma_k\leq \frac{b_2}{\log 2}
  \gamma_k,~~~~~~~~{\rm with}~~~~~~~~
  \gamma_k = {\rm max}|\varphi_k|
\end{equation}
which is proven in Appendix C, and due to (\ref{Cinequality})
\begin{equation}
  \bigg|\int \eta_0 u'_k d^2z\bigg|<b_2'\gamma_k~.
\end{equation}
Then we have directly
\begin{equation}\label{mubound}
  |\mu_{k}|=\bigg|-\frac{\int\eta_0 u'_{k-1} d^2z}{\int\psi d^2z}\bigg|<
  b'_2~~\frac{1}{\lambda\int_{R_c}\eta_0d^2z}\gamma_{k-1} \equiv b'\gamma_{k-1}
\end{equation}
and putting together
\begin{equation}\label{varphibound}
  |\varphi_k|=\bigg|\frac{\eta_0}{\zeta_1}u'_{k-1}+\mu_k\frac{\psi}{\zeta_1}
  \bigg|<\frac{\eta_0}{\eta}
  Z b_2\gamma_{k-1}+\alpha' b'\gamma_{k-1}<(mb_2+\alpha'b')\gamma_{k-1}
\end{equation}
where
\begin{equation}
\alpha'={\rm max}\bigg|\frac{\psi}{\zeta_1}\bigg|
\end{equation}
i.e.
\begin{equation}
  \gamma_k<(m b_2+\alpha' b')\gamma_{k-1}
\end{equation}
and thus convergence for 
\begin{equation}\label{implicitineq}
 \lambda_1~(mb_2+\alpha'b')<1~.
\end{equation}
We have however to keep in mind that while $b_2$ and $m$,
see eq.(\ref{mdefinition}),  are independent of $\lambda$, $\alpha'$ and
$b'$ depend on the original $\lambda$ and we need to prove that for
small enough $\lambda$ the inequality (\ref{implicitineq}) holds for
$\lambda_1=\lambda$.

In computing $\alpha'$ we have the ratio
\begin{equation}
\frac{\psi}{\zeta_1}=\frac{\lambda \eta_0\nu^h - \Delta \nu^h}{\zeta_1}~.
\end{equation}
As $\psi$ vanishes for $|\zeta|<R'$ we shall examine the two contributions
separately for $|\zeta|>R'$.
With regard to the first term $\lambda \eta_0\nu^h $ we have
\begin{equation}\label{l0bound}
  \lambda\frac{\eta_0}{\zeta_1} \nu^h\leq\lambda\frac{\eta_0}{\zeta_1}\leq
  \lambda m L(z-z_P)<\lambda m l_0
\end{equation}
with $l_0$ the value of $L(z-z_P)$ on the circle of radius $R'$, being the
first inequality a consequence of $h<0$ and $\nu\geq1$ and the second
a consequence of the fact that $L$ decreases for $|z-z_P|$ increasing.

The second term gives the contribution
\begin{equation}
-\frac{h\nu^{h-1}\Delta\nu+h(h-1)\nu^{h-2}\nabla\nu\cdot\nabla\nu}{\zeta_1}~.
\end{equation}
Such term vanishes for $|\zeta|>R$ and thus taking into account that
$h(h-1)=\lambda t$, $h<0$, we have that for $\lambda t<1$ the above
ratio, is in absolute value less than $\lambda b_3$, $b_3$ being a new
constant.

Summing up we reach for $\lambda<1/t$
\begin{equation}
  \alpha'<\lambda (m l_0+b_3)
\end{equation}
and going back to (\ref{varphibound}) and (\ref{mubound})
\begin{equation}
\gamma_k<\bigg(b'_2+\frac{m l_0+b_3}{\int_{R_c}\eta_0 d^2z}\bigg)\gamma_{k-1}
\end{equation}
and thus we have convergence for $\lambda_1=\lambda<1/t$ and
\begin{equation}
\lambda\bigg(b'_2+\frac{m l_0+b_3}{\int_{R_c}\eta_0 d^2z}\bigg)<1~.
\end{equation}

We recall furthermore that
\begin{equation}
|u_k|<\frac{b_2}{L(z-z_P)}\gamma_k
\end{equation}
with $L(z-z_P)>\log 2$  
which assures the uniform convergence of the $u_k$. It is not difficult
using the inequalities of Appendix B and Appendix C to prove that we can
exchange the sum with the Laplacian $\Delta$ and thus we have reached
the solution of (\ref{simplerequation}).

\bigskip

Then by a finite number of extension steps we can reach the solution of
\begin{equation}\label{beta1equation2}
\Delta u=\eta_0 u-\beta_1~.
\end{equation}
From the solution of (\ref{beta1equation2}) we obtain the solution of
\begin{equation}\label{lambdauprimeeq}
\Delta u=\eta u-\beta_1~~~{\rm with}~~~~~\eta=\eta_0+\eta'~.
\end{equation}
In fact we can write
\begin{equation}\label{lambdauprimelambda1}
\Delta u=(\eta_0+\lambda_1\eta')u-\beta_1
\end{equation}
and we have
\begin{equation}
r=\frac{\eta'}{\eta_0}=\frac{(\eta(\zeta,z_P)-\eta_E(\zeta))\rho(\zeta)
}{\eta_E(\zeta)
\rho(\zeta)+\eta(\zeta,z_P)(1-\rho(\zeta))}~.
\end{equation}
This ratio is different from $0$ only in support of $\rho$.
As $\eta$ and $\eta_E$ are both positive we have
\begin{equation}
|r|=\frac{|\eta(\zeta,z_P)-\eta_E(\zeta)|\rho(\zeta)}{\eta_E(\zeta)
  \rho(\zeta)+\eta(\zeta,z_P)(1-\rho(\zeta))}<
\frac{|\eta(\zeta,z_P)-\eta_E(\zeta)|}{\eta_E(\zeta)}
\Theta_\rho(\zeta)
\end{equation}
where $\Theta_\rho$ is the characteristic function of the support of
$\rho$ i.e. $|\zeta|<R$. As already noticed, see
eq.(\ref{etaprimeovereta0}), for $R$ sufficiently small the maximum of
$|r|$ is less than $1$.  Then the expansion in $\lambda_1$ of
eq.(\ref{lambdauprimelambda1}) converges uniformly for $\lambda_1=1$ and we
have solved eq.(\ref{beta1equation}).

\section{More than one parabolic singularity}\label{twoparabolic}

We consider explicitly the extension to the case of two parabolic
singularities.

Similarly to what done in the previous section we write
$\eta=\eta_0+\eta'$ with 
\begin{equation}
\eta_0 = \eta_{E_1}\rho_1+\eta_{E_2}\rho_2+ \eta(1-\rho_1-\rho_2)
\end{equation}
where
\begin{equation}
  \eta_{E_1}=\frac{t_1}{|z-z_{P_1}|^2\log^2|z-z_{P_1}|^2},~~~~
  \eta_{E_2}=\frac{t_2}{|z-z_{P_2}|^2\log^2|z-z_{P_2}|^2}
\end{equation}
and consider the simpler equation, where $\eta$ has been substituted by
$\eta_0$
\begin{equation}\label{simpler2}
\Delta u = \lambda \eta_0u-\beta_{12}  
\end{equation}
where $\beta_{12}$ has the tamed singularities at $z_{P_1}$ and
$z_{P_2}$ due to the subtraction procedure described by
(\ref{subtraction},\ref{beta1equation}) in the previous section.
Write
\begin{equation}\label{decomposition2}
u=u'+\mu_1\nu_1^{h}+\mu_2\nu_2^{h}
\end{equation}
with
\begin{equation}
  \nu_1=\bigg(c_{11}\log^{a_1}\frac{1}{|z-z_{P_1}|^2}-1\bigg)~\rho_1(z-z_{P_1}) +1
+\bigg(c_{12}\log^{a_2}\frac{1}{|z-z_{P_2}|^2}-1\bigg)\rho_2(z-z_{P_2})
\end{equation}
and
\begin{equation}
    \nu_2=\bigg(c_{21}\log^{a_1}\frac{1}{|z-z_{P_1}|^2}-1\bigg)~\rho_1(z-z_{P_1}) +1
+\bigg(c_{22}\log^{a_2}\frac{1}{|z-z_{P_2}|^2}-1\bigg)\rho_2(z-z_{P_2})
\end{equation}
where $c_{ij}>0$, $a_j>0$ and we use $R_j<1/\sqrt{e}$, with
\begin{eqnarray}
&& 4 ha_1(ha_1-1)=\lambda~t_1\\
&& 4 ha_2(ha_2-1)=\lambda~t_2~.
\end{eqnarray}
The functions $\nu_j$ are strictly positive:
$\nu_j>\min(c_{j1},c_{j2},1)\equiv m_j$.  We can set $a_1=1$ and
determine $a_2$ as a function of $h$. For small $h$, $a_2$ goes over
to $t_2/t_1$.  Now we have two parameters $\mu_1$ and $\mu_2$ at our
disposal, with which we can satisfy the two conditions of finiteness
at infinity and of the vanishing of the solution $u'$ at $P_2$, while
the vanishing at $P_1$ is assured by the use of the subtracted Green
functions $G_1$.  Thus eq.(\ref{simpler2}) becomes
\begin{equation}
\Delta u'=\lambda\eta_0 u'+\mu_1\psi_1+\mu_2\psi_2-\beta_{12}
\end{equation}
with
\begin{equation}\label{thepsis}
  \psi_1= \lambda \eta_0 \nu_1^{h}-\Delta \nu_1^{h}~,~~~~~~~
  \psi_2= \lambda \eta_0 \nu_2^{h}-\Delta \nu_2^{h}~.
\end{equation}

We remark that the functions $\psi_i$ vanish exactly in the
neighborhoods $|z-z_{P_1}|<R_1'$ and $|z-z_{P2}|<R'_2$.

We have that
\begin{equation}
  \int\psi_1d^2z=\lambda\int\eta_0 \nu_1^{h}d^2z>
  \lambda\int_{R_{c}}\eta_0 d^2z>0~.
\end{equation}
being $R_{c}$ the region
$(|z-z_{P{_1}}|>R_1)\cap(|z-z_{P_2}|>R_2)$, where $\nu_1$ equals
$1$.

Similarly
\begin{equation}
  \int\psi_2d^2z=\lambda\int\eta_0 \nu_2^{h}d^2z>
  \lambda\int_{R_{c}}\eta_0 d^2z>0~.
\end{equation}

Again given a $\lambda$, on which $\psi_1$ and $\psi_2$ depend, see
eq.(\ref{thepsis}), we shall consider the equation
\begin{equation}\label{twoparablambda1}
  \Delta u'=\lambda_1\eta_0~u'+\mu_1(\lambda_1)\psi_1+
  \mu_2(\lambda_1)\psi_2-\beta_{12}
\end{equation}
and expand in $\lambda_1$.  At the order zero we have the equation
\begin{equation}
  \Delta u'_{0}=\mu_{10}\psi_1+\mu_{20}
\psi_2-\beta_{12}\equiv\zeta_{12}\varphi_0
\end{equation}
where $\zeta_{12}$ has been defined in analogy with eq.(\ref{zeta1}) as
\begin{equation}
  \zeta_{12}=\frac{\eta}{Z L_{12}},~~~~{\rm with}~~~~L_{12}(z)=L(z-z_{P_1})+
  L(z-z_{P_2})
\end{equation}
and at the order $k$, $k>0$ we have the equation
\begin{equation}\label{k2equation}
  \Delta u'_{k}=\eta_0 u'_{k-1}+\mu_{1k}\psi_1+
  \mu_{2k}\psi_2\equiv \zeta_{12}\varphi_k~.
\end{equation}
The $\mu_{1k}$ and $\mu_{2k}$ have to be chosen as to satisfy the regularity
of $u_k$ at infinity  i.e.
\begin{equation}\label{infinity2eq}
\mu_{1k}\int\psi_1d^2z+\mu_{2k}\int\psi_2d^2z+\int\eta_0 u'_{k-1}d^2z=0
\end{equation}
and the vanishing of $u'_k$ at $z_{P_2}$ i.e. 
\begin{eqnarray}\label{P2vanishing}
  &&\mu_{1k}\int G_1(z_{P_2},z')\psi_1(z')d^2z'+ \mu_{2k}\int
  G_1(z_{P_2},z')\psi_2(z')d^2z'\nonumber\\
  &&+\int G_1(z_{P_2},z')\eta_0(z')u'_{k-1}(z')d^2z'=0
\end{eqnarray}
where $G_1(z,z')$ is the Green function subtracted at $z_{P_1}$, which
assures the vanishing of $u'_k$ at $z_{P_1}$.

The equations determining the $\mu_{ik}$ are
\begin{equation}\label{musystem}
  \begin{pmatrix}
    m_{11}&m_{12}\\
    m_{21}&m_{22}
    \end{pmatrix}
  \begin{pmatrix}
    \mu_{1k}\\
    \mu_{2k}
  \end{pmatrix}=
  \begin{pmatrix}
    n_{1k}\\
    n_{2k}
    \end{pmatrix}~.
\end{equation}
It is important that the matrix $m_{ij}$ does not depend on the order $k$
of the expansion.

In the limit of small $\lambda$ we have 
$m_{11}=m_{12}=\lambda\int \eta_0 d^2z\equiv \lambda c>0$.
\begin{equation}\label{m21eq}
m_{21}= \int G_1(z_{P_2},z')\psi_1(z')d^2z'
\end{equation}
where the integral is convergent due to the vanishing of $\psi_1$ in the
neighborhood of the $z_{P_1},z_{P_2}$ while $m_{22}$ is obtained replacing
$\psi_1$ with $\psi_2$ in eq.(\ref{m21eq}).

In the limit $h\rightarrow 0$ we go over to
\begin{equation}
  m_{21}= h \int_{R'_c} G_1(z_{P_2},z')\bigg(-\frac{4}{t_1}\eta_0-
  \nu_1^{-1}\Delta\nu_1
  +\nu_1^{-2}\nabla\nu_1\cdot\nabla\nu_1\bigg)d^2z'
\end{equation}
where the integration is restricted to the region $R'_c=(|z-z_{P_1}|>R'_1)\cap
(|z-z_{P_2}|>R'_2)$.
With regard to the $n_{ik}$ terms the main point is that
\begin{equation}\label{b12bound}
  |u'_k|<b_{12} \frac{1}{L_{12}} \gamma_k
\end{equation}
as a consequence of the bound of Appendix C. In fact as $u'_k$
vanishes at $z_{P_2}$ it can be written as
\begin{equation}
  u'_k=\frac{1}{4\pi}\int\log\bigg|\frac{z-z'}{z_{P_2}-z'}\bigg|^2
  \zeta_1(z')\varphi_k(z') d^2z'
\end{equation}
and then we can apply the original argument of Appendix C.

For $n_{1k}$ we have due to (\ref{b12bound})
\begin{equation}\label{n1k}
  |n_{1k}|=\bigg|-\int \eta_0 u'_{k-1}d^2z\bigg|< b'_{12}
    \gamma_{k-1}
\end{equation}
while for $n_{2k}$ we have
\begin{eqnarray}\label{n2k}
  &&|n_{2k}|=\bigg|-\int G_1(z_{P_2},z')\eta_0(z')u'_{k-1}(z')d^2z'\bigg|\leq
  \frac{b_{12}\gamma_{k-1}}{4\pi}\int\bigg|\log|z_{P_2}-z'|^2\bigg|
  \frac{\eta_0(z')}{L_{12}(z')}d^2z'\nonumber\\
  &&+\frac{b_{12}\gamma_{k-1}}{4\pi}\int\bigg|\log|z_{P_1}-z'|^2\bigg|
  \frac{\eta_0(z')}{L_{12}(z')}d^2z'
\end{eqnarray}
where both integrals are convergent.

\bigskip

The solution of the two equations (\ref{musystem}) is
\begin{equation}
   \begin{pmatrix}
    \mu_{1k}\\
    \mu_{2k}
  \end{pmatrix}
= \frac{1}{\lambda c(m_{22}-m_{21})}
   \begin{pmatrix}
    m_{22}&-\lambda c\\
    -m_{21}&\lambda c
   \end{pmatrix}
      \begin{pmatrix}
    n_{1k}\\
    n_{2k}
  \end{pmatrix}
\end{equation}
with $n_{jk}$ given by the last terms in
eqs.(\ref{infinity2eq},\ref{P2vanishing}), bounded by
(\ref{n1k},\ref{n2k}) and thus for small $\lambda$
\begin{eqnarray}\label{varphikbound}
  &&\mu_{1k}\approx\frac{m_{22} n_{1k}}{\lambda c
    (m_{22}-m_{21})}\equiv \frac{c_{1k}}{\lambda}\nonumber\\
  &&\mu_{2k}\approx\frac{-m_{21} n_{1k}}{\lambda c
    (m_{22}-m_{21})}\equiv \frac{c_{2k}}{\lambda}
\end{eqnarray}
where due to eq.(\ref{n1k}) we have
\begin{equation}
|c_{1k}|< p_{1}\gamma_{k-1},~~~~~~~~|c_{2k}|< p_{2}\gamma_{k-1}~.
\end{equation}  

Going back to eq.(\ref{k2equation}) we have

\begin{equation}\label{Blambda}
  |\varphi_k|< b_{12}\frac{\eta_0}{\zeta_{12}}\frac{\gamma_{k-1}}{L_{12}}
  +|\mu_{1k}|\bigg|\frac{\psi_1}{\zeta_{12}}\bigg|
  +|\mu_{2k}|\bigg|\frac{\psi_2}{\zeta_{12}}\bigg|<
  (b_{12}  m +\frac{p_{1}}{\lambda}\alpha_1'
  +\frac{p_{2}}{\lambda}\alpha_2')\gamma_{k-1}
  \equiv B(\lambda)\gamma_{k-1}
\end{equation}
where $\alpha'_j=\max|\psi_j/\zeta_{12}|$.  We shall show that $B(0)$
is finite.  To compute the $\alpha'_j$ of eq.(\ref{Blambda}) one
uses the fact that
\begin{equation}
\psi_1=\lambda\eta_0\nu_1^{h}-\Delta \nu_1^{h}=
\lambda\eta_0\nu_1^{h}-
h\nu_1^{h-1}\Delta\nu_1-h(h-1)\nu_1^{h-2}\nabla\nu_1\cdot\nabla\nu_1.
\end{equation}

Then for $|z-z_{P_1}|<R_1'$ or $|z-z_{P_2}|<R_2'$ we have zero. For
$R_1'<|z-z_{P1}|<R_1$ or $R_2'<|z-z_{P1}|<R_2$ we have for $0<-h<1$
\begin{equation}
  \bigg|\frac{\psi_1}{\zeta_{12}}\bigg|
  \leq \frac{\lambda}{m_1}\max\frac{\eta_0}{\zeta_{12}}+\frac{(-h)}{m_1^2}
  \max\bigg|\frac{\Delta\nu_1}{\zeta_{12}}\bigg|+
  \frac{(-h)(1-h)}{m_1^3}\max\frac{\nabla\nu_1\cdot\nabla\nu_1}{\zeta_{12}}
\end{equation}
where the maxima refer to the intervals $R'_1<|z-z_{P_1}|<R_1$ and
$R'_2<|z-z_{P_2}|<R_2$.  For $(|z-z_{P_1}|>R_1)\cap(|z-z_{P_2}|>R_1)$
we have $\nu_1=1$, $\eta_0=\eta$ and thus
\begin{equation}
  \bigg|\frac{\psi_1}{\zeta_{12}}\bigg|\leq\lambda Z L_{12}<
  \lambda~(l_{10}+l_{20})
\end{equation}
where $l_{j0}$ are defined as in eq.(\ref{l0bound}). The same argument
holds for $\psi_2/\zeta_{12}$.

Thus $B(0)$ is finite. This proves that the expansion in $\lambda$ has
a non zero convergence radius.  In fact for $\lambda$ less than some
proper $\lambda_0$ we shall have $B(\lambda)<B(0)+\varepsilon$.  Going
back to eq.(\ref{twoparablambda1}) we have that the series in
$\lambda_1$ converges uniformly for
\begin{equation}
|\lambda_1| B(\lambda)<1~.
\end{equation}
Choose $\lambda <\lambda_0$ and at the same time
$\lambda(B(0)+\varepsilon)<1$~.

Then we have convergence of the $\gamma_k$ for $\lambda_1=\lambda$ and
using the inequality
\begin{equation}
  |u'_{k}|=\bigg|\int G_1(z,z')\zeta_{12}(z')\varphi_k(z')d^2z\bigg|<
  b_{12} \frac{\gamma_k}{L_{12}}
\end{equation} 
we have the uniform convergence of the $u'$.  Thus we have solved the
equation (\ref{simpler2}) for $\lambda$ sufficiently small.  The extension
procedure to the solution of eq.(\ref{simpler2}) with $\lambda=1$ is
the same as in the one parabolic singularity case and the same for the
extension to the original equation where $\eta$ instead of $\eta_0$
appears.

\section{The inheritance of the analytic properties}\label{inheritance}

In the present section we prove, using the series expansions developed
previously, that the conformal factor $\phi$ of
eq.(\ref{liouvilleequation}) depends in real analytic way on the
positions of the singularities $z_k,z_P$. We shall consider explicitly
the case when only one parabolic singularity is present, being the
general case a straightforward extension of this one.  Obviously it is
sufficient to look at the case when only one singularity moves and we
shall examine explicitly the case when the moving singularity is the
parabolic one i.e. $z_P$; this is the most difficult case.

We need the detailed structure of the most important function which
appears in the iteration procedure i.e. of $\theta\equiv\beta r$. We
are interested in the problem when $z_P$ varies in a domain $D_P$
around a $z^0_P$, say $|z_P-z^0_P|<R_P$ which excludes all others
singularities.  We choose $R_P$ equal to $1/4$ the minimal distance of
$z^0_P$ from the singularities $z_k$.  We know that
$0<r_1<\frac{\theta}{\beta}<r_2<\infty$ where the bounds $r_1$ and
$r_2$ can be taken independent of $z_P$ for $z_P\in D_P$.  For the
function $\beta(z,z_P)$ we point out the uniform bound for $|z|>2
~{\max} (|z_k|,|z_P|)$ and $z_P\in D_P$
\begin{equation}\label{generalbound}
  \beta<\frac{c}{(1+z\bar z)^2}~.
\end{equation}
The function $\theta(z,z_P)$ is explicitly given by
\begin{equation}
  \theta(z,z_P) = ((z-z_P)(\bar z-\bar z_P))^{-1}
  \prod_k ((z-z_k)(\bar z-\bar z_k))^{-2\eta_k}
   e^{I(z,z_P)}
\end{equation}
where
\begin{equation}\label{I}
I(z,z_P)=\frac{1}{4\pi}\int\log|z-z'|^2 \beta(z',z_P)d^2z'~.
\end{equation}
The properties of $I(z,z_P)$ both in $z$ and $z_P$ are worked out in
Appendix D. As done in that appendix it is useful to introduce for
$\zeta\equiv z-z_P$, $|\zeta|<R_1$, the function $\hat
I(\zeta,z_P)\equiv I(\zeta+z_P,z_P)$, where $R_1$ is such that the
disk of radius $R_1$ includes only the $z_P$ singularity.  Similarly
one defines the function $\hat\theta(\zeta,z_P)$. The result is that
$\hat\theta(\zeta,z_P)$ is analytic in $z_P$ for $z_P\in D_P$ and in
$\zeta$ for $\zeta\neq 0$. It is also useful to introduce an other
region defined by $|z-z_P|>R_2$ for some $R_2$, $0<R_2<R_1$. In such a
region $\theta(z,z_P)$ is analytic in $z_P$, $z_P\in D_P$, and in $z$
except at the singularities $z_k$.

Actually in the process of the previous sections we met with several
cases in which functions equivalent to $\theta$ appear, i.e. function
$\eta$ such that $0<r_1<\frac{\eta}{\theta}<r_2<\infty$ for $z_P\in
D_P$ thus we shall generically deal with such functions $\eta(z,z_P)$.

The transformation which plays the master role in the present
developments is the solution of eq.(\ref{recursiveeq}) which vanishes
at $z=z_P$ i.e. 
\begin{equation}\label{resolvent}
u'_k(z,z_P)=\int G_1(z,z')\zeta_1(z',z_P)\varphi_k(z',z_P)d^2z'~,
\end{equation}
where $G_1$ is the subtracted Green function
\begin{equation}
G_1(z,z')=\frac{1}{4\pi}\log\bigg|\frac{z-z'}{z_P-z'}\bigg|^2
\end{equation}
and the source has the property
\begin{equation}\label{sumrule3}
\int\zeta_1(z',z_P)\varphi_k(z',z_P)d^2z'=0~.
\end{equation}
where the function $\zeta_1$ is given in (\ref{zeta1}).

We are interested in the analytic properties of the transformation
(\ref{resolvent}) and in order to do that it is useful again, as done
in Appendix D and in \cite{PMelliptic}, to define in the region
$|\zeta|=|z-z_P|\leq R_1$ for a function $f(z,z_P)$, $\hat
f(\zeta,z_P)\equiv f(\zeta+z_P,z_P)$ and thus write
\begin{eqnarray}\label{resolvent1}
&&\hat u'_k(\zeta,z_P)=\int_{{\cal R}_1}G_1(\zeta+z_P,\zeta'+z_P)|^2
  \hat\zeta_1(\zeta',z_P)\hat \varphi_k
  (\zeta',z_P)d^2\zeta'+\nonumber\\
&&\int_{{\cal R}_{1c}}G_1(\zeta+z_P,z')\zeta_1(z',z_P)\varphi_k(z',z_P) d^2z'~.
\end{eqnarray}
where ${\cal R}_1$ is the region $|\zeta|<R_1$ and ${\cal R}_{1c}$ its
complement. It is also useful to consider a $R_2$ with 
$0<R_2<R_1$ and write for $|z-z_P|>R_2$
\begin{eqnarray}\label{resolvent2}
&&u'_k(z,z_P)=\int_{{\cal R}_2}G_1(z,z_P+\zeta')
  \hat\zeta_1(\zeta',z_P)
  \hat\varphi_k(\zeta',z_P)d^2\zeta'+\nonumber
  \\ &&\int_{{\cal R}_{2c}}G_1(z,z')\zeta_1(z',z_P)\varphi_k(z',z_P))d^2z'~.
\end{eqnarray}
The great advantage of such a decomposition is that in taking the
derivative w.r.t. $z_P$ only absolutely integrable terms appear. This
is paid by the appearance in the derivative of a contour term due to
the dependence on $z_P$ of the integration regions ${\cal R}_{1c}$ and
${\cal R}_{2c}$ in the second part of
eqs.(\ref{resolvent1},\ref{resolvent2}).  This contour term is very easily
dealt with as done after eq.(\ref{contourderivative}).

\bigskip

We shall now show that some boundedness and real analyticity
properties of the source $\varphi_k(z,z_P)$ are inherited by
$u_k(z,z_P)$ through the transformation (\ref{resolvent}).  We shall
always work with $z_P\in D_P$ where $D_P$ was described at the
beginning of the present section and does not contain any
singularity $z_k$.  The real analyticity is proven by showing the
existence the complex derivatives w.r.t. $z$ and $\bar z$ or w.r.t
$z_P$ and $\bar z_P$.  Due to the symmetry of the problem it is
sufficient to prove analyticity w.r.t. $z$ and $z_P$.

We give now to the main theorem of this section

\bigskip

\underline{Theorem 6.1}

\bigskip

Suppose that $\varphi_k$ has the following properties for $z_P\in D_P$

\smallskip

P1. $\varphi_k(z,z_P)$ is bounded

P2. $\hat\varphi_k(\zeta,z_P)$ is analytic in $\zeta$ for
$|\zeta|>0$, $|\zeta|\neq R',R$

P3. $\hat\varphi_k(\zeta,z_P)$ for $z_P\in D_P$, $0<|\zeta|<R_1$ is
analytic in $z_P$ with $\frac{\partial\hat\varphi_k(\zeta,z_P)}
{\partial z_P}$ bounded

P4. $\varphi_k(z,z_P)$ is analytic in $z$ for $|z-z_P|>R_2$, $z=\infty$
included, except at $z=z_k$

P5. $\varphi_k(z,z_P)$ is analytic in $z_P$ with $\frac{\partial
  \varphi_k(z,z_P)}{\partial z_P}$ bounded for $z_P\in D_P$,
$|z-z_P|>R_2$

\bigskip

then $u'_k$ given by eq.(\ref{recursiveeq}) i.e.
\begin{equation}
  u'_k(z,z_P)=\int G_1(z,z')\zeta_1(z',z_P)\varphi_k(z',z_P)d^2z'
\end{equation}

satisfy the same properties
P1-P5.

\bigskip

\bigskip

\underline{Proof}

\bigskip

The inheritance of property P1 is a consequence of the inequality
proven in the Appendix C.

We shall first consider analyticity in $\zeta$ and $z$ i.e.
properties P2 and P4.
The rules for computing the derivatives of
eq.(\ref{resolvent1},\ref{resolvent2}) are summarized in Appendix F.

In the coordinates $\zeta,z_P$ the $\hat\theta$, from which all the
$\eta$ and $\zeta_1$ follow, has the following structure
\begin{equation}\label{thetahat}
  \hat\theta(\zeta,z_P) = (\zeta\bar\zeta)^{-2} \prod_{k}
  (|\zeta+z_P-z_k|^2)^{-2\eta_k} e^{\hat I(\zeta,z_P)},~~~~~~~~|\zeta |<R_1~.
\end{equation}  

The derivative of first term of eq.(\ref{resolvent1})
w.r.t. $\zeta$ is given by
\begin{equation}
  \frac{1}{4\pi}\int_{{\cal R}_{1}}
  \frac{1}{\zeta-\zeta'}\hat\zeta_1(\zeta',z_P)\hat\varphi_k(\zeta')d^2\zeta'
\end{equation}
where taking the derivative under the integral sign is legal for
$|\zeta|\neq(0, R', R)$ due
to rule F1 of Appendix F and the absolute integrability of
\begin{equation}
\hat\zeta_1(\zeta',z_P)\hat\varphi_k(\zeta')
\end{equation}
on ${{\cal R}_{1}}$.
The derivative of the second term is
\begin{equation}
  \frac{1}{4\pi}\int_{{\cal R}_{1c}}
  \frac{1}{\zeta-z'+z_P}\zeta_1(\zeta',z_P)\varphi_k(\zeta')d^2z'
\end{equation}
again obtained applying rule F1 as on ${{\cal R}_{1c}}$, $\zeta+z_P-z'$
is lower bounded and the behavior $\zeta_1\sim 1/(z\bar z)^2$
at infinity assures absolute integrability.

As for P4  i.e. analyticity in $z$ of (\ref{resolvent2}) the derivative of
the first piece gives
\begin{equation}
  \frac{1}{4\pi}\int_{{\cal R}_{2}}
  \frac{1}{z-\zeta'-z_P}\hat\zeta_1(\zeta',z_P)\hat\varphi_k(\zeta')d^2\zeta'
\end{equation}
again applying rule F1.
The derivative of the second piece gives for $z\neq z_k$
\begin{equation}
  \frac{1}{4\pi}\int_{{\cal R}_{2c}}
  \frac{1}{z-z'}\zeta_1(z',z_P)\varphi_k(z')d^2z'
\end{equation}
where taking the derivative under the integral sign is legal due to
rule F1 and the behavior at infinity,
$\zeta_1\sim (z\bar z)^{-2}$ of $\zeta_1$ to satisfy F2.

\bigskip

We come now to the analyticity in $z_P$ i.e. properties P3 and P5.
With respect to the first term in (\ref{resolvent1}), the derivative w.r.t.
$z_P$ is
\begin{equation}
  \frac{1}{4\pi}\int_{{\cal R}_{1}}
  \log\bigg|\frac{\zeta-\zeta'}{\zeta'}\bigg|^2
  \frac{\partial(\zeta_1(z',z_P)\varphi_k(z',z_P))}
  {\partial z_P}d^2\zeta'~.
\end{equation}
In computing the derivative the condition F3 given in
Appendix F for taking the derivative under the integral sign is
satisfied as $\hat\varphi_k$ and $\frac{\partial
  \hat\varphi_k}{\partial z_P}$ are bounded in ${\cal R}_1 \times
D_P$, i.e. properties P1 and P3. In fact the product term in
eq.(\ref{thetahat}) satisfy properties P1, P2, P3.

In taking the derivative w.r.t. $z_P$ of the second term in
(\ref{resolvent1}) we must take into account that the
integration region ${\cal R}_{1c}$ moves as $z_P$ varies. Then the
derivative of the second integral appearing in (\ref{resolvent1}) is
\begin{eqnarray}\label{contourderivative}
&& \frac{1}{4\pi}\int_{{\cal R}_{1c}} \frac{\partial}{\partial z_P}
  \bigg[\log\bigg|\frac{\zeta+z_P-z'}{z_P-z'}\bigg|^2
    \zeta_1(z',z_P)\varphi_k(z',z_P)\bigg]
  d^2z'\nonumber\\ &-& \frac{1}{8\pi i}\oint_{\partial {\cal R}_{1c}}
  \log\bigg|\frac{\zeta+z_P-z'}{z_P-z'}\bigg|^2 \zeta_1(z',z_P)\varphi_k(z',z_P)
  d\bar z'~.
\end{eqnarray}
The logarithms in the above equation are not singular for
$\zeta\in{\cal R}_1$ which makes the differentiation under the
integral sign legal as satisfying condition F3 of Appendix F.

\bigskip

We come now to the $u'_k(z,z_P)$ with $|z-z_P|>R_2$ where we use
expression (\ref{resolvent2}). We have for the first term
\begin{eqnarray}
&& \frac{1}{4\pi}\int_{{\cal R}_{2}}
  \frac{1}{z_P+\zeta'-z}
  \hat\zeta_1(\zeta',z_P)\hat\varphi_k(\zeta',z_P)
  d^2\zeta'+\nonumber\\
  &&\frac{1}{4\pi}\int_{{\cal
      R}_{2}}\log\bigg|\frac{z-z_P-\zeta'}{\zeta'}\bigg|^2
  \frac{\partial}{\partial
    z_P}\big(\hat\zeta_1(\zeta',z_P)\hat\varphi_k(\zeta',z_P)\big) d\zeta'~.
\end{eqnarray}
Again the first term is justified by
\begin{equation}
   \int_{{\cal R}_{2}} \hat\zeta_1(\zeta',z_P)|\varphi_k(\zeta',z_P))|d^2\zeta'
<\infty
\end{equation}
and the second by property P3 and integrability of $\hat\zeta_1$ and
$\frac{\partial \zeta_1}{\partial z_P}$ on $R_2$.

For the derivative of the second term in (\ref{resolvent2}) we have
\begin{eqnarray}
&& \int_{{\cal R}_{2c}} G_1(z,z')\frac{\partial}{\partial
    z_P}(\zeta_1(z',z_P)\varphi_k(z',z_P))
  d^2z'+\nonumber\\
  &&-\frac{1}{4\pi}\int_{{\cal R}_{2c}}
  \frac{1}{z_P-z'}\zeta_1(z',z_P)\varphi_k(z'z_P)d^2z'+\nonumber\\
  &&-\frac{1}{2i}\oint_{\partial{\cal
      R}_{2c}} G_1(z,z')\zeta_1(z'z_P)\varphi_k(z',z_P)d\bar z'~.
\end{eqnarray}

\bigskip

Here the second contributions is due to the dependence of $G_1$ on
$z_P$ and the last is the contour integral due to the motion of
${\cal  R}_{2c}$ as $z_P$ varies where the single terms are justified
by the properties P1, P5 of $\varphi_k$.

\bigskip

We are left to examine the neighborhood of $z=\infty$ which, with the
behavior (\ref{betainfinity}) for the $\beta$ at infinity and the
consequent behavior of the $\theta$ and $\zeta_1$ is a regular point.

In fact with $\tilde u'_k(x,z_P)\equiv u'_k(1/x,z_P)$ and
$\tilde\theta(y,z_P)\equiv\theta(1/y,z_P)$ and using (\ref{resolvent})
and condition (\ref{sumrule3}) we have
\begin{equation}\label{tildeequation}
\tilde
u'_k(x,z_P)=\frac{1}{4\pi}\int\log\bigg|\frac{x-y}{\frac{1}{z_P}-y}\bigg|^2
  \frac{\tilde\zeta_1(y,z_P)}
  {(y\bar y)^2}\tilde\varphi_k(y,z_P)d^2y~.
\end{equation}
Exploiting the analyticity of $\tilde\zeta_1(y,z_P)/(y\bar y)^2$ for
$|y|<1/\Omega$ we have that (\ref{tildeequation}) has complex
derivative w.r.t. $x$ for $|x|<1/\Omega$ thus proving the analyticity
in $x$ of $\tilde u'_k$ around $x=0$.

From the previous eq.(\ref{tildeequation}) we see that $\tilde
u'_k(x,z_P)$ is analytic in the polydisk $|x|<1/\Omega$ and $z_P\in
D_P$. This assures not only that $u_k$ at infinity is bounded, a result
that we knew already from the treatment of sections
\ref{poincareprocedure}, \ref{linearsection} and of Appendix C
but that $\frac{\partial u'_k}{\partial z_P}$ is uniformly bounded for
all $z$ with $|z-z_P|>R_2$. Thus we have reproduced for $u'_k$ the
properties P1-P5 completing the proof of the theorem. \hfill $\Box$

\bigskip

We extend now by induction the result to all the $u'_k$ and to their
sums.

\bigskip

\underline{Theorem 6.2}

The properties P1-P5 of the $u'_k$ hold for all $k$ and for
the sums of sections \ref{linearsection}, \ref{twoparabolic} and
\ref{poincareprocedure}.

\underline{Proof}

To complete the induction procedure we have to examine the properties
of $\varphi_k$ given by
\begin{equation}
  \zeta_1\varphi_k = \eta u'_{k-1} +\mu_k\psi =
  \eta u'_{k-1} +\mu_k(\lambda\eta \nu^h-\Delta\nu^h)~.
\end{equation}
With regard to the first term $\eta u'_{k-1}/\zeta_1$ we notice that the
$\eta/\zeta_1= Z L$ and thus we have analyticity in $z,z_P$ for
$z_P\in D_P$ and $z\neq(z_k,z_P), |z-z_P|\neq R',R$. With regard to
the $\psi$ term we recall that $\mu_k$ is analytic in $z_P$ for
$z_P\in D_P$ and $\psi$ it vanishes for $|z-z_P|<R'$. Thus $\varphi_k$
satisfies the properties P1-P5.

We have now to extend the properties P1-P5 of the $u'_k$ to their sum.

To establish the real analyticity of the sum of the series we shall
exploit the well known result \cite{ahlfors} that given a sequence of
analytic functions $f_n$ defined in a domain $\Omega$ which converge
to $f$ uniformly on every compact subset of $\Omega$ then their sum is
analytic in $\Omega$ and the series of the derivatives $f'_n$ converge
uniformly to $f'$ on every compact subset of $\Omega$. Such result was
already applied to the elliptic case in \cite{PMelliptic}.

In the process of extending the range of the parameter $\lambda$
which intervene in the various extension steps one repeats exactly the
procedure described above. We recall that the extension steps
necessary for achieving such extensions are always finite in number.
Then following the simple procedure of \cite{PMelliptic} given the
initial domain of $z_P$ i.e. $|z_P-z_{P_0}|<R_P$ we shall have
analyticity of the conformal factor in $z$ for $|z-z_P|\neq (R',R)$
and $z\neq (z_k,z_P)$ and analyticity in $z_P$ for
$|z_P-z_{P_0}|<R_P-\varepsilon$ for any $\varepsilon>0$. \hfill $\Box$

\bigskip

Up to now in this section we have been concerned with the analytic
properties of the solution of the linear inhomogeneous equation
studied in section \ref{linearsection}. We saw in section
\ref{poincareprocedure} that the original non linear equation is
reduced to the solution of the infinite system (\ref{nonlinearsystem})
of such linear inhomogeneous equations. To each of them we apply the
above described procedure to establish the analytic property of their
solutions and due to the uniform convergence we have that such
properties extend to their sum.

In the expansion scheme described in section \ref{poincareprocedure}
the following modification of the function $\theta$ appears in the
second and subsequent steps
\begin{equation}
\theta \rightarrow \theta e^u
\end{equation}  
where $u$ is the solution of the preceding step in the procedure, see
eq.(\ref{secondstep}). As proven in the present section $u$ is bounded
and analytic in $z$ and $z_P$ for $z\neq(z_k,z_P)$, $|z-z_P|\neq
(R',R)$ and $z_P\in D_P$ and then such analytic properties are shared
by $\theta e^u$.

Due to the freedom in choosing the values of $R'$ and $R$ and the uniqueness
theorem \cite{lichtenstein,PMexistence} we have
real analyticity in $z$ everywhere except at the sources $z_k,z_P$
and real analyticity in $z_P$ for $|z_P-z_{P_0}|<R_P-\varepsilon$.

\bigskip

Finally we recall that the conformal field $\phi(z,z_P)$ is given in
terms of $u$ by
\begin{equation}
  \phi(z,z_P)=u(z,z_P)+\nu(z,z_P)=u(z,z_P)-\log|z-z_P|^2
  -2\sum_k\eta_k\log|z-z_k|^2+
I(z,z_P)
\end{equation}  
where the analytic properties of $I(z,z_P)$ have already been given in
the Appendix D. Thus we conclude that $\phi(z,z_P)$ is
real analytic in $z_P$ and in $z$ for $|z_P-z_P^0|< R_P-\varepsilon$
and for $z\neq z_k,z_P$.

\section{The real analyticity of the accessory parameters}
\label{parametersanalyticity}

In the previous section we proved that the conformal factor
$\phi(z,z_P)$ is a real analytic function of $z$ for $z\neq (z_k,
z_P)$ and that for $z\neq(z_k,z_P)$ it is a real analytic function
of $z_P$ for $z_P\in D_P$.

The real analyticity
of the accessory parameters is a consequence of this fact.

\bigskip

\underline{Theorem 7.1}

The accessory parameters are real analytic functions of the source
positions.

\bigskip

\underline{Proof}

The auxiliary differential equation is
\begin{equation}
y''(z)+Q(z)y(z)=0
\end{equation}  
where $Q(z)$ is expressed in terms of the conformal factor $\phi$ as
\begin{eqnarray}
&&Q(z)=-e^{\frac{\phi}{2}}\frac{\partial^2 }{\partial
    z^2}e^{-\frac{\phi}{2}}=\nonumber\\
&& \sum_k\frac{\eta_k(1-\eta_k)}{(z-z_k)^2}+
  \sum_k\frac{b_k}{2(z-z_k)}+\nonumber\\
&& \sum_P\frac{1}{4(z-z_P)^2}+
  \sum_P\frac{b_P}{2(z-z_P)}~.
\end{eqnarray}  

Consider a singularity $z_k$ or $z_P$, different from the moving
singularity, and a circle $C_k$ around it, of radius such that no other
singularity is contained in it. Given the conformal factor
$\phi=u+\nu$ we have that both $u$ and $\nu$ are real analytic
functions of $z$ and $z_P$ for $z$ in an annulus containing $C_k$ and
$z_P\in D_P$. The accessory parameter $b_k$ can be expressed in terms
of $\phi$ as
\begin{equation}\label{contourintegral}
b_k=\frac{1}{i\pi}\oint_{C_k} Q(z) dz~.
\end{equation}  
Due to the analyticity of $\phi$ we can associate to any point of
$C_k$ a polydisk $D_z\times D_P$ where the $\phi$ is real
analytic. Due to the compactness of $C_k$ we can extract a finite
covering provided by such polydisks.  It follows then that the
integral (\ref{contourintegral}) is a real analytic function of $z_P$
for $z_P\in D_P$.  Thus we have that all the accessory parameters
$b_k$ and $b_P$, except the one relative to the moving singularity,
are real analytic functions of the position of the moving source.
With respect to the accessory parameter relative to the moving
singularity we recall that due to the Fuchs relations
\cite{menottiHigherGenus} it is given in terms of the others $b_k,b_P$
and thus also this last accessory parameter is a real analytic
function of the position of the moving source. The reasoning obviously
holds for the dependence on the position of any singularity keeping
the others fixed, thus concluding the proof of the real analyticity on
all source positions.\hfill $\Box$

\bigskip

We have been working explicitly in the case of the sphere topology
with an arbitrary number of elliptic and parabolic singularities. This
treatment goes beyond the results of
\cite{menottiAccessory,menottiHigherGenus,menottiPreprint} where it
was found that in the case of the sphere with four sources we had real
analyticity almost everywhere. With the almost everywhere attribute we
could not exclude the occurrence of a number of cusps in the
dependence of the accessory parameters on the position of the
sources. Here we proved real analyticity everywhere and for any number
of sources and thus the occurrence of cusps is excluded.  Obviously
the whole reasoning holds when the positions of the singularities are
all distinct.  What happens when two singularities meet has been
studied only in special cases in \cite{kra} and \cite{HS1,HS2}.

With regard to higher genus the treatment extends immediately to the
torus topology with any number of elliptic and parabolic singularities
obviously satisfying the topological Picard inequalities. In the paper
\cite{PMelliptic} we gave a detailed treatment of the case of the
torus with elliptic singularities showing that the accessory
parameters are analytic functions not only of the positions of the
sources but also of the modulus of the torus. Along the same lines one
extends the treatment also in presence of parabolic singularities. The
simplifying feature of the torus case is that we know for it the
explicit form of the Green function along with its analytic
properties.

For higher genus we do not possess the explicit form
of the Green function and we have a representation of the analogue of
the Weierstrass $\wp$ function only for genus 2 \cite{komori}. Thus
one should employ more general arguments for the analyticity of the
Green function and for the expression of the $\beta$ function.
For $g>1$ the best approach appears to be
the use of the representation of the Riemann surface using the
fuchsian domains in the upper half-plane. For carrying through the
program, in absence of explicit forms of the Green function, one
should establish its analytic dependence on the moduli and also one
should provide an analytic $\beta$ satisfying the correct boundary
conditions.

\section{Discussion and conclusions}\label{conclusions}

In the paper \cite{PMelliptic} it was proved that on the sphere
topology in presence of any number of elliptic singularities, the
accessory parameters are real analytic functions of the source
positions. The result was also given in the case of the torus
with any number of elliptic singularities.

In the present paper we extended such a result to the case when in
addition of elliptic singularity any number of parabolic
singularities is present.

Such an extension is not trivial due to the highly singular character
of the parabolic sources. In fact the typical integral which intervenes
in the solution of the Liouville equation
\begin{equation}
\frac{1}{4\pi}\int\log|z-z'|^2 f(z')d^2z'  
\end{equation}
in presence of a parabolic singularity i.e.
\begin{equation}
f(z')\sim\frac{1}{|z'-z_P|^2 \log^2|z'-z_P|^2}  
\end{equation}
diverges for $z\rightarrow z_P$. This makes a straightforward use
of the methods of \cite{PMelliptic} inapplicable.

Due to this fact in the process it is necessary to introduce in an
intermediate step a regulator \cite{poincare}. We choose a $C^\infty$
regulator even though as shown in Appendix F, with some work it
is possible also to adopt a real analytic regulator.

The result extends immediately to the torus topology where the
accessory parameter depend both on the source positions and on the
modulus of the torus.

The same continuation method should work also for higher genus
i.e. $g\geq 2$.  Here the main problem is the knowledge of the
analytic properties of the Green functions.

The present paper completes the program started in \cite{kra} and
pursued with the method of analytic varieties in
\cite{menottiAccessory, menottiPreprint}. The method applied here and
in \cite{PMelliptic} i.e. the Le Roy-Poincar\'e continuation method of
\cite{leroy,poincare} revealed itself particularly powerful as
it applies to the most general situation of general elliptic
singularities and parabolic singularities.

For higher genus i.e. $g>1$ the best approach appears to be the use of
the representation of the Riemann surface using the fuchsian domains
in the upper half-plane. For carrying through the program, in absence
of explicit forms of the Green function, one should establish its
analytic dependence on the moduli and also one should provide an
analytic $\beta$ satisfying the correct boundary conditions.

\eject

\section*{Appendix A}

In this appendix we give the analytic form of the functions which
appear in our treatment of the problem. The choice of such functions
is not unique; it is a matter of convenience. On the other hand
the uniqueness theorem \cite{lichtenstein,PMexistence}
assures us that the final results are
independent of such a choice.

1. The $\beta$ function. This is the most important function as from
it the $\theta$ function which appears in the fundamental equation is
expressed. We recall that such function should be positive with well
defined singularities or zeros at the elliptic points, well defined
singularities at the parabolic points and behavior $1/(z\bar z)^2$ at
infinity. Furthermore the $\beta$ has to satisfy the sum rule
(\ref{sumrule}).

For simplicity we give the case with three parabolic singularities
and any number of elliptic singularities. We start from the function
\begin{eqnarray}\label{trial}
  &&\prod
  (|z-z_k|^2)^{-2\eta_k}\times\nonumber\\ &&
  \frac{(1+z\bar z)^{2\sigma-1}}
{|z-z_{P_1}|^2 L^2(z-z_{P_1}) |z-z_{P_2}|^2 L^2(z-z_{P_2}) |z-z_{P_3}|^2 L^2(z-z_{P_3})}
    \times\nonumber\\
  && \bigg(A_1 |z-z_{P_2}|^2|z-z_{P_3}|^2
  +A_2|z-z_{P_1}|^2|z-z_{P_3}|^2+ A_3|z-z_{P_1}|^2|z-z_{P_2}|^2\bigg)
\end{eqnarray}

with $\sigma=\sum\eta_k$ and where $L$ is given by
\begin{equation}
L(\zeta)= \log\frac{1+2|\zeta|^2}{|\zeta|^2}~.
\end{equation}
Such an expression is positive has the correct behavior at infinity
and we fix the $A_i$ to have the coefficient $8$ at the parabolic
singularities. The $A_j$ are real analytic functions of the position of
the singularities. 
We have the topological restriction $\sum_k 2\eta_k+\sum_P 1 >2(1-g)$
where $g$ is the genus of the surface and $g=0$ as we work on the
sphere.

In order to satisfy the sum rule 
\begin{equation}\label{sumrule2}
\frac{1}{4\pi}\int \beta d^2z= \sum 2\eta_k +N_P -2=2\sigma+N_P-2
\end{equation}
if the integral of (\ref{trial}) is less or equal to the l.h.s. of
(\ref{sumrule2}) we multiply by the factor
\begin{equation}\label{normfactor}
  \frac{1+N |z-z_{P_1}|^2|z-z_{P_2}|^2|z-z_{P_3}|^2}
  {1+|z-z_{P_1}|^2|z-z_{P_2}|^2|z-z_{P_3}|^2}~.
\end{equation}
and fix $N$ as to satisfy (\ref{sumrule2}).  Such factor does not
alter the positivity the asymptotic behavior and the coefficient at
the parabolic singularities. $N$ is a real analytic function of the
position of the singularities \cite{PMelliptic}.  If the integral of
(\ref{trial}) is larger or than the l.h.s. of (\ref{sumrule2}) we use
the reciprocal of (\ref{normfactor}).

\bigskip

2. We give now an explicit expression of the function $v$, which by a
process of subtraction was used to tame the parabolic singularities of
$\beta$.

\begin{eqnarray}
&&  v= \frac{|z-z_{P_1}|^2 |z-z_{P_2}|^2 A_3}{(|z-z_{P_1}|^2+1)
    (|z-z_{P_2}|^2+1)}+\nonumber\\
&&  \frac{|z-z_{P_1}|^2 |z-z_{P_3}|^2 A_2}{(|z-z_{P_1}|^2+1)
    (|z-z_{P_3}|^2+1)}+\nonumber\\
&&  \frac{|z-z_{P_2}|^2 |z-z_{P_3}|^2 A_1}{(|z-z_{P_2}|^2+1)
    (|z-z_{P_3}|^2+1)}~.
\end{eqnarray}

The $A_j$ are chosen as to cancel the parabolic singularities of
$\beta$ through the mechanism of eq.(\ref{beta1equation}) and they are
real analytic functions of the $z_k,z_P$.  The contribution of $\Delta
v$ at infinity goes to zero.

\section*{Appendix B}
In \cite{PMelliptic} it was proven that in presence of elliptic
singularities if
\begin{equation}
  \int\eta(z')\varphi(z')d^2z'=0
\end{equation}
the following bound holds
\begin{equation}
  \bigg|\frac{1}{4\pi}\int \log|z-z'|^2\eta(z')\varphi(z')d^2z'\bigg|<
  b\max|\varphi|~.
\end{equation}
Such a result holds also in presence of tamed parabolic singularities
i.e. if
\begin{equation}
  \int\zeta_1(z')\varphi(z')d^2z'=0
\end{equation}
the following bound holds
\begin{equation}
  \bigg|\frac{1}{4\pi}\int \log|z-z'|^2\zeta_1(z')\varphi(z')d^2z'\bigg|<
  b_1 \max|\varphi|
\end{equation}
with $b_1$ independent of $z_P$ for $z_P\in D_P$.  The proof goes
along the same method used in \cite{PMelliptic} i.e.  isolating the
contribution of the integral over non overlapping disks of radius
$a<\frac{1}{4}$ and also less than $1/4$ of the minimal distance of
$z_P$ from the other singularities, and a remainder.  Thus we have
to examine the contribution of a disk of radius $a$ around a
tamed parabolic singularity
\begin{equation}
\frac{1}{4\pi}\int_{D_a} \log|z-z'|^2\zeta_1(z')\varphi(z')d^2z'~.
\end{equation}
We have simply to examine it for $|z|<2\Omega$ where $\Omega$ is the
radius of a circle enclosing all singularities.
Using $\zeta=z-z_P$, for $|\zeta|>2 a$ we have
\begin{equation}
  \bigg|\frac{1}{4\pi}\int_a \log|z-z'|^2\zeta_1(z')\varphi(z')d^2z'\bigg|<
  \frac{1}{4\pi}\bigg(|\log|\zeta|^2|+ 2\log 2\bigg)
  \int_a\zeta_1(z')dz'\max|\varphi|
\end{equation}
and as we are working for $|z|<2\Omega$ it gives rise to a finite bound.
For $a<|\zeta|<2a$ explicit integration gives the bound
\begin{equation}
\frac{M}{8}\frac{\log\frac{1}{|\zeta|^2}}{\log^2 a^2}\max|\varphi|
\end{equation}
where $M$ is such that
\begin{equation}
\zeta_1<\frac{M}{\zeta^2|\log|\zeta|^2|^3}
\end{equation}
in the disk of radius $a$. Finally for $|\zeta|<a$ explicit computation
gives the bound
\begin{equation}\label{Mbound}
\frac{M}{4}\bigg(-\frac{1}{\log
  a^2}+\frac{1}{2\log|\zeta|^2}\bigg)\max|\varphi|
\end{equation}
where in (\ref{Mbound}) both logarithms are negative.

\section*{Appendix C}

In this appendix we prove a bound which will be fundamental in the
developments of the paper. Consider the equation

\begin{equation}\label{simpleeq}
\Delta u(z) =\zeta_1(z)\varphi(z)
\end{equation}
with, see eq.(\ref{zeta1})
\begin{equation}
  \zeta_1(z)=\frac{\eta(z)}{Z(z) L(z-z_P)}~,
\end{equation}
\begin{equation}
L(z-z_P) =\log\frac{(1+2|z-z_P|^2)}{|z-z_P|^2}
\end{equation}
and
\begin{equation}
  \int\zeta_1(z)\varphi(z)d^2z=0
\end{equation}
which assures the boundedness of the solution at infinity.  The
solution of (\ref{simpleeq}) which vanishes for $z=z_P$ is given by
\begin{equation}
\int G_1(z,z') \zeta_1(z')\varphi(z')dz'
\end{equation}
where
\begin{equation}
G_1(z,z') =\frac{1}{4\pi}\log\bigg|\frac{z-z'}{z_P-z'}\bigg|^2
\end{equation}

\underline{Lemma C.1}

Given
\begin{equation}\label{basicintegral}
A=\int  G_1(z,z') \zeta_1(z')\varphi(z')d^2z'
\end{equation}
with
\begin{equation}\label{restriction2}
  \int\zeta_1(z')\varphi(z')d^2z'=0
\end{equation}

we have
\begin{equation}\label{secondPoincare}
  |A|<\frac{b_2}{L(z-z_B)}~\max|\varphi|
\end{equation}
with $b_2$ independent of $z_P$ for $z_P\in D_P$.

\bigskip

\underline{Proof}

We shall use $z-z_P=\zeta$. Moreover we shall consider a disk $D_a$
around $z_P$ of radius $a$ less than $1/4$ of the minimal distance of
$z_P$ from all the other singularities; we shall also choose
$a<1/4$.

Under restriction (\ref{restriction2}) we have due to the results of
the previous Appendix B, 
$|A|<2b_1 \max|\varphi|$ and thus for $|z-z_P|>\frac{a}{2}$ we can write
\begin{equation}\label{fromPoncare1}
|A|< 2 b_1\frac{L(\frac{a}{2})}{L(\zeta)} \max|\varphi|
\end{equation}
being $L(\zeta)$ a decreasing function of $|\zeta|=|z-z_P|$.  We now
write $A$ as the sum of an integral over the disk $D_a$ of radius $a$
and its complement.  With regard to the contribution of the complement
$D_{ac}$ of $D_a$, for $|\zeta|<a/2$ we have
\begin{eqnarray}
&&\frac{1}{4\pi}\bigg|\int_{D_{ac}}
\log\bigg|1-\frac{\zeta}{\zeta'}\bigg|^2
\zeta_1(\zeta')\varphi(\zeta')d^2\zeta'\bigg|\leq 
\frac{1}{4\pi}\int_{D_{ac}}
  \bigg|\log\bigg|1-\frac{\zeta}{\zeta'}\bigg|^2\bigg|
  \zeta_1(\zeta')|\varphi(\zeta')|d^2\zeta'<\nonumber\\
&&  \frac{\max|\varphi|}{\pi}|\zeta|\int_{D_{ac}}\frac{\zeta_1(\zeta')}
       {|\zeta'|}d\zeta'
\end{eqnarray}
which comply the bound (\ref{secondPoincare}). Thus we are left
with the contribution of $D_a$ for $|\zeta|<a/2$.  Such contribution
is computed explicitly.
Using $r=|\zeta|$ and $\rho=|\zeta'|$ we have the bound
\begin{equation}
  M \frac{\max |\varphi|}{4\pi}\int_{D_a} \bigg|
  \log(r^2+\rho^2-2r\rho\cos\theta)-
  \log\rho^2\bigg| \frac{1}{\rho^2|\log^3\rho^2|}\rho d\rho d\theta
\end{equation}
where $M$ is such that for $z\in D_a$ we have
\begin{equation}
 \zeta_1 < \frac{M}{\rho^2|\log \rho^2|^3}~.
\end{equation}
The expression
\begin{equation}
  \log(r^2+\rho^2 - 2 r \rho\cos\theta)-\log\rho^2
\end{equation}
is not of definite sign. We can majorize the positive part i.e. the
contribution of the region where
\begin{equation}
r^2-2 r\rho \cos\varphi>0
\end{equation}

by
\begin{equation}
\log(r+\rho)^2-\log\rho^2=2\log\bigg(1+\frac{r}{\rho}\bigg)~.
\end{equation}

For $r\log\frac{1}{r}\equiv c<a$, we split the integral, from $0$ to
$c$ and from $c$ to $a$. We have
\begin{eqnarray}\label{bound1}
&& -2\int_c^a
  \log(1+\frac{r}{\rho})\frac{1}{\log^3\rho}\frac{d\rho}{\rho}<
  -2\log(1+\frac{r}{c})\bigg[\frac{l^{-2}}{-2}\bigg]^{\log
    a}_{\log c}<\nonumber\\
  &&\frac{r}{c}(\frac{1}{(\log
    a)^2}-\frac{1}{(\log
    c)^2})<\frac{1}{\log\frac{1}{r}}\frac{1}{\log^2a}~.
\end{eqnarray}

For the integral from $0$ to $c=r\log\frac{1}{r}$ we use, for $x>0$
\begin{equation}
  \log(1+x)\leq \log 2+|\log x|
\end{equation}
and we have
\begin{equation}\label{bound5}
-2\int_0^{r\log\frac{1}{r}}(\log 2+|\log r - \log
\rho|)\frac{1}{\log^3\rho}\frac{d\rho}{\rho }<
-2\int_0^{r\log\frac{1}{r}}(\log 2+\log\frac{1}{r} - \log
\rho)\frac{1}{\log^3\rho}\frac{d\rho}{\rho }
\end{equation}
The $\log 2$ term gives
\begin{equation}\label{log2term}
\frac{\log 2}{\log^2(r\log\frac{1}{r})}
\end{equation}
while the term $\log\frac{1}{r}$ gives
\begin{equation}\label{log1overrterm}
\frac{\log\frac{1}{r}}{\log^2(r\log\frac{1}{r})}~.
\end{equation}
For $r<1$ we have
\begin{equation}\label{remark}
  \frac{1}{2}\log \frac{1}{r} >\log(\log\frac{1}{r})>0 
\end{equation}
and thus both (\ref{log2term}) and (\ref{log1overrterm}) are majorized by
$\frac{\rm const}{\log\frac{1}{r}}$.

Finally
\begin{equation}\label{bound2}
  2\int_0^{r\log\frac{1}{r}}\frac{1}{\log^2\rho}
  \frac{d\rho}{\rho}=-\frac{2}{\log(r\log\frac{1}{r})}
  <\frac{4}{\log\frac{1}{r}} 
\end{equation}
again due to (\ref{remark}).

We come now to modulus of the negative part.

The negative part can be majorized by
\begin{eqnarray}
&& \int d\theta \int^a_{\frac{r}{2\cos\theta}} (\log\rho^2 - \log(\rho^2 + r^2
  - 2 r\rho \cos\theta))p(\rho)\rho d\rho <\nonumber\\
  && \int d\theta\int^a_{\frac{r}{2\cos\theta}}
  (\log\rho^2 - \log(\rho -r)^2)p(\rho)\rho d\rho
\end{eqnarray}
with
\begin{equation}
p(\rho) = \frac{1}{-\rho^2 \log^3 \rho^2}>0~.
\end{equation} 
As for $\rho>r/2$ the integrand stays positive we have the
majorization
\begin{eqnarray}
&&  \int d\theta
  \int^a_{\frac{r}{2\cos\theta}} (\log\rho^2 - \log(\rho-r)^2)p(\rho)
  \rho d\rho<\nonumber\\
&&2\pi\int^a_{\frac{r}{2}} (\log\rho^2 - \log(\rho-r)^2)p(\rho)\rho d\rho=
2\pi\int_{\frac{r}{2}}^a[-\log(1-\frac{r}{\rho})^2]
  \frac{1}{-\log^3\rho^2}\frac{d\rho}{\rho}~.
\end{eqnarray}

We notice that the maximum of $\frac{1}{|\log\rho^2|^3}$ is reached at
$\rho=a$ as we are working for $0<\rho<a<1/4$. Splitting the integral
from $r/2$ to $r\log\frac{1}{r}$ and from $r\log\frac{1}{r}$ to $a$ we
have for the second part, using $-\log r>1/2$ and $|\log(1-x)^2|<4 x$
for $0<x<\frac{1}{2}$
\begin{eqnarray}\label{bound3}
&&  \int_{r\log\frac{1}{r}}^a
    -\log(1-\frac{r}{\rho})^2\frac{1}{|\log\rho^2|^3}
  \frac{d\rho}{\rho}<
\frac{4 r}{|\log a^2|^3}\int_{r\log\frac{1}{r}}^a \frac{d\rho}{\rho^2}=\nonumber\\
&&  \frac{4}{|\log a^2|^3}\bigg(-\frac{r}{a}+\frac{1}{\log\frac{1}{r}}\bigg)
<\frac{4}{|\log a^2|^3}\frac{1}{\log\frac{1}{r}}~.
\end{eqnarray}
We are left with
\begin{eqnarray}
&& \int^{r\log\frac{1}{r}}_{\frac{r}{2}}
  [-\log(1-\frac{r}{\rho})^2]\frac{1}{(-\log\rho^2)^3}
  \frac{d\rho}{\rho}< \frac{1}{8|\log (r\log\frac{1}{r})|^3}\times
  \int^{\log\frac{1}{r}}_{\frac{1}{2}}[-\log(1-\frac{1}{y})^2]
  \frac{dy}{y}\nonumber\\ &&<
  \frac{1}{8|\log (r\log\frac{1}{r})|^3}\times
  \int^{\infty}_{\frac{1}{2}}[-\log(1-\frac{1}{y})^2]\frac{dy}{y}
  =\frac{{\rm const}}{8|\log(r\log\frac{1}{r})|^3}~.
\end{eqnarray}
But we have seen already, eq.(\ref{remark}), that for $r<1$ we have
\begin{equation}
\big|\log(r(\log\frac{1}{r}))\big|>\frac{1}{2}\log\frac{1}{r}
\end{equation}
and thus the bound
\begin{equation}\label{bound4}
\frac{{\rm const}}{\log^3\frac{1}{r}}
\end{equation}
which is stronger than the bound (\ref{secondPoincare}).  Combining
the inequalities
(\ref{bound1},\ref{bound5},\ref{bound3}) we reach the
bound (\ref{secondPoincare}). \hfill $\Box$

\bigskip

\section*{Appendix D}

In the present appendix we prove the properties of the integral
$I(z,z_P)$ which are used in the text. As described in the text, it is
useful to consider a disk of radius $R_1$ around the parabolic
singularity $z_P$ which contains no other singularity and also a disk
of radius $0<R_2<R_1$ again with center $z_P$. 

\bigskip

\underline{Lemma D1}

\smallskip

The integral
\begin{equation}
I(z,z_P)=  \frac{1}{4\pi}\int \log|z-z'|^2\beta(z',z_P) d^2z'
\end{equation}
is everywhere finite and continuous in $z$ except at $z=z_P$, analytic
in $z$ for $z\neq z_k,z_P$. $\hat I(\zeta,z_P)\equiv I(\zeta+z_P,z_P)$ is
analytic in $z_P$ for $z_P\in D_P$ for $0<|\zeta|<R_1$ while $I(z,z_P)$
is analytic in $z_P$ for $|z-z_P|>R_2$.  At infinity $I(z,z_P)$
behaves like $(\sum_k 2\eta_k+ \sum_P 1-2)\log|z|^2 $ and we have
$|\frac{\partial I(z,z_4)}{\partial z_P}|< {\rm const}~\log|z|^2$.

\bigskip

\underline{Proof}

The finiteness is due to the convergence of the integral and the continuity
for $z\neq z_k$ easily follows. 

The analyticity in $z$ is proven by splitting the integral in a disk
$c$ around a point $z_0$ and a remainder and using the fact that
the integral
\begin{equation}
\frac{1}{4\pi}\int_c \log|z-z'|^2 s(z')d^2z'
\end{equation}
is analytic in $z$ at $z_0$ if $s(z)$ is analytic in $c$, applying rule
F1 of Appendix F and for the contribution of the complement we apply
rule F2.

The analyticity of $\hat I(\zeta,z_P)$ in $z_P$ is obtained as done in
\cite{PMelliptic} by splitting the integral in two parts the first for
$|\zeta'|<R_1$ and the other for $|z'-z_P|>R_1$. The complex
derivative w.r.t. $z_P$ is given by three terms: the derivative of the
integrand of the first and second integral and a contour term due to
the motion in the second integral of the $|z'-z_P|>R_1$ region.
Taking the derivative under the integral sign in the first and second
integral is legal as the expressions can be majorized by a function
of $z'$, independent of $z_P$ for $z_P\in D_P$.

The analyticity of $I(z,z_P)$ in $z_P$ for $|z-z_P|> R_2$ is
again proven by splitting the integration domain in a region
$|\zeta'|<R_2$ and $|z'-z_P|>R_2$.

The three terms expressing the derivative are bound by
$c_1\log(z\bar z+1)+c_2$ for $|z|>2 \Omega$ where $\Omega$ is the
radius of a circle which encloses all singularities and $c_1$, $c_2$
are positive constants. \hfill  $\Box$

\section*{Appendix E}
As we discussed in detail in the text, due to the highly singular
nature of the parabolic sources at an intermediate stage it is
necessary to introduce a regulator, which is later removed.

In the text we used the $C^\infty$ regulators $\rho$. As discussed in
section \ref{linearsection} the outcome is that we prove
analyticity of $u(z,z_P)$ except at the sources and on the circles
$|z-z_P|=R',R$.

Due to the freedom in choosing $R',R$ and the uniqueness theorem at
the end we have analyticity of $u(z,z_P)$ for all $z$ except
at the sources.

In this appendix we show that it is possible, even though a little more
complicated, to work only with analytic regulators.

To achieve that we replace the $\rho(\zeta)$ of section
\ref{linearsection} with
\begin{equation}
e^{-\frac{r^4}{R^4}}
\end{equation}
where $r=|z-z_P|$. $R$ will be taken less than $1$. Then
using the same definition of $\nu$ we have
\begin{equation}
\nu = 1+(-\log r^2-1)\rho >c>0~.
\end{equation}
We give again a decomposition of the type $\eta=\eta_0+\eta'$
with
\begin{equation}
\eta_0 = \eta_N\rho+(1-\rho)\eta
\end{equation}
where 
\begin{equation}
  \eta_N=\frac{t}{r^2 L^2(r)}\frac{Z(z)}{Z(z_P)}
  ~~~~~~{\rm with}~~~~~~~~Z(z)=\prod_s\frac{(|z-z_s|^2)^{-2\eta_s}}
      {(1+|z-z_s|^2)^{-2\eta_s}}
\end{equation}
where $s$ runs on the $\eta_s<0$.
\begin{equation}
\eta' = \eta-\eta_0=\rho(\eta-\eta_N)~.
\end{equation}
We cannot use here the $\eta_E$ of section \ref{linearsection}
because it would introduce a spurious singularity at $r=1$.
The ratio
\begin{equation}
\frac{\eta'}{\eta_0}
\end{equation}
for $R$ sufficiently small is in absolute value less than $1$.  
For $\zeta_1$  we use as in section \ref{linearsection}
\begin{equation}
\zeta_1 =\frac{\eta}{Z L}~.
\end{equation}

Again the crucial point in the development is the proof that
\begin{equation}
\alpha'<{\rm const}~h
\end{equation}
where
\begin{equation}
\alpha' = \max\bigg|\frac{\psi}{\zeta_1}\bigg|
\end{equation}
with
\begin{equation}
\psi = \lambda\eta_0\nu^h-\Delta\nu^h
\end{equation}

For $r$ larger than some $r_0$ we have
\begin{equation}
\bigg|\frac{4h(h-1)}{t} \frac{\eta_0}{\zeta_1}\nu^h- \frac{h\nu^{h-1}\Delta\nu
  +h(h-1)\nu^{h-2}\nabla\nu\cdot\nabla\nu}{\zeta_1}\bigg|<{\rm const} ~h
\end{equation}

Then all the problem is what happens at small $r$ i.e. $0<r<r_0$ and,
as discussed at the end of section \ref{twoparabolic}, for small $h$.

From explicit computation we find
\begin{equation}
  \bigg|\frac{\psi}{\zeta_1}\bigg| =h~ O(r\log r)
\end{equation}
which is bounded for $r<r_0$. We can then follow the argument given at
the end of section \ref{twoparabolic} to prove that for sufficiently
small $\lambda$ we have the convergence of the series for $u'$.

\section*{Appendix F}

As explained in section \ref{lichtenstein} for simplicity we use the
complex notation $z=x+iy$, $\bar z=x-i y$.

In several places in the text we need to prove the analyticity of
certain integral expressions of the type
\begin{equation}\label{simpleintegral2}
\frac{1}{4\pi}\int\log|z-z'|^2 f(z',z_P) d^2z'
\end{equation}
The standard procedure is the following:

1. Analyticity in $z$.

Consider a disk of radius $a$ around a point $z_0$ and write
(\ref{simpleintegral2}) as the sum of the integral extended to the
disk $D$ and its complement $D_c$.

If, Condition F1, $D$ is contained in a domain of analyticity of
$f(z,z_P)$ then the first contribution is analytic in $z$ for $z$
belonging to the interior of $D$ and its derivative is given by
\begin{equation}\label{Dderivative}
  \frac{1}{4\pi}\int_D\frac{1}{z-z'} f(z',z_P) d^2z'
\end{equation}
The contribution of the second integral is analytic for $z$ belonging
to the interior of $D$ and its derivative is given by
\begin{equation}\label{derintegral}
  \frac{1}{4\pi}\int_{D_c}\frac{1}{z-z'} f(z',z_P) d^2z'
\end{equation}
provided  the integrand in eq.(\ref{derintegral}), is majorized
by a function $g(z',z_P)$ independent of $z$  i.e.
\begin{equation}
\bigg|\frac{1}{z_1-z'} f(z',z_P)\bigg|<g(z'z_P)
\end{equation}
for $z_1$ in a neighborhood of $z$ i.e. for $|z_1-z|<\varepsilon$ and $g$
is absolutely integrable.
We shall refer this as Condition F2.

A simple sufficient condition is that, with $\delta$ the distance of
$z$ from the boundary i.e. $\delta=a-|z-z_0|$, we have
\begin{equation}
\int_{D_c}\frac{|f(z',z_P)|}{|z'-z_0|-a+\frac{\delta}{2}} d^2z'<\infty
\end{equation}
Thus if Condition F1 and Condition F2 are satisfied, then the integral
(\ref{simpleintegral2}) is analytic in $z$ and its derivative is given
by (\ref{Dderivative}) with $D$ replaced by the whole plane.

\bigskip

2. Analyticity in $z_P$.

If $f(z,z_P)$ is analytic in $z_P$ the integral is analytic in $z_P$
and its derivative given by
\begin{equation}
  \frac{1}{4\pi}\int\log|z-z'|^2 \frac{\partial f(z',z_P)}
  {\partial z_P}d^2z'
\end{equation}
if in a neighborhood of $z_P$ the following bound holds
\begin{equation}
\bigg|\log|z-z'|^2 \frac{\partial f(z',z_P)}
  {\partial z_P}\bigg|<g(z')
\end{equation}
with $g(z')$ absolutely integrable. We shall refer to
this condition as Condition F3.

\eject

\vfill


\end{document}